\documentclass[twoside]{article}

\usepackage[accepted]{aistats2025}
\usepackage[commenters={T}]{shortex} 
\RequirePackage{amsthm,amsmath,amsfonts,amssymb}
\usepackage[round]{natbib}

\usepackage{tikz}
\usetikzlibrary{positioning}

\usetikzlibrary{shapes,decorations,arrows,calc,arrows.meta,fit,positioning}
\tikzset{
    -Latex,auto,node distance =1 cm and 1 cm,semithick,
    state/.style ={circle, draw, minimum width = 0.75cm},
    point/.style = {circle, draw, inner sep=0.04cm,fill,node contents={}},
    bidirected/.style={Latex-Latex,dashed},
    el/.style = {inner sep=2pt, align=left, sloped}
}

\definecolor{julia1}{rgb}{0, 0.60, 0.98} 
\definecolor{julia2}{rgb}{0.89, 0.43, 0.28}
\definecolor{julia3}{rgb}{0.24, 0.64, 0.30}
\definecolor{julia4}{rgb}{0.76, 0.44, 0.82}
\definecolor{julia5}{rgb}{0.67, 0.55, 0.09}
\definecolor{julia6}{HTML}{00aaae}
\definecolor{julia7}{HTML}{ed5e93}
\definecolor{julia8}{HTML}{c68225}

\newcommand{\DUGS}{K_\text{DUGS}}

\newcommand{\kappar}{\kappa_\mathrm{r}}

\newcommand{\kappac}{\kappa_\mathrm{cor}}
\newcommand{\ESS}{T_\mathrm{eff}}

\newcommand\ndatasetstotal{12}
\newcommand\ndatasetswin{11}

\begin{document}

\fancyhead[CO]{\small\bfseries Luu, Xu, Surjanovic, Biron-Lattes, Campbell, Bouchard-C\^{o}t\'{e}}

\twocolumn[
  \aistatstitle{Is Gibbs sampling faster than Hamiltonian Monte Carlo on GLMs?}
  \aistatsaddress{
    \setlength\tabcolsep{0pt}
    \begin{tabular*}{0.95\linewidth}{@{\extracolsep{\fill}} ccc }
    \bf Son Luu & \bf Zuheng Xu & \bf Nikola Surjanovic \\
    University of British Columbia & University of British Columbia & University of British Columbia \\
    & & \\
    \bf Miguel Biron-Lattes & \bf Trevor Campbell & \bf Alexandre Bouchard-C\^{o}t\'{e} \\
    University of British Columbia & University of British Columbia & University of British Columbia
    \end{tabular*}
  }
]

\begin{abstract}
  The Hamiltonian Monte Carlo (HMC) algorithm is 
  often lauded for its ability to effectively sample from high-dimensional distributions. 
  In this paper we challenge the presumed domination of HMC for the Bayesian analysis of GLMs. 
  By utilizing the structure of the compute graph rather than the graphical model, we show a 
  reduction of the time per sweep of a full-scan Gibbs sampler from $O(d^2)$ to $O(d)$, 
  where $d$ is the number of GLM parameters. 
  A simple change to the implementation of the Gibbs sampler allows us 
  to perform Bayesian inference on high-dimensional GLMs that are 
  practically infeasible with traditional Gibbs sampler implementations. 
  We empirically demonstrate a substantial increase in effective sample size per time 
  when comparing our Gibbs algorithms to state-of-the-art HMC algorithms.   
  While Gibbs is superior in terms of dimension scaling, 
  neither Gibbs nor HMC dominate the other: we provide numerical 
  and theoretical evidence that 
  HMC retains an edge in certain circumstances thanks to its advantageous condition number scaling. 
  Interestingly, for GLMs of fixed data size, we observe that increasing dimensionality 
  can stabilize or even decrease condition number, shedding light on the empirical advantage 
  of our efficient Gibbs sampler. 
\end{abstract}

\section{INTRODUCTION}
\label{sec:intro}

\begin{figure}[!t]
	\centering
	\scalebox{0.8}{
	\begin{tikzpicture}
		\node[state, fill = julia2] (1) {$+$};
		\node[state, fill = julia2] [below = of 1, yshift = 0.5cm] (p) {$\mu_i$};
		\node[state] [below right = of 1] (Y) {$Y_i$};
		\node[state, fill = julia2] [below = of p, yshift = 0.5cm] (l) {$\ell_i$};
		\node[state, fill = julia2] (22) [above = of 1, yshift = -0.5cm] {$\times$};
		\node[state] (21) [left = of 22, xshift = -0.5cm] {$\times$};
		\node[state] (23) [right = of 22, xshift = 0.5cm] {$\times$};
		\node[state] (31) [above = of 21, xshift = -0.55cm, yshift = -0.5cm] {$\theta_1$};
		\node[state] (32) [above = of 21, xshift = 0.55cm, yshift = -0.5cm] {$x_{i1}$};
		\node[state, fill = julia2] (33) [above = of 22, xshift = -0.55cm, yshift = -0.5cm] {$\theta_2'$};
		\node[state] (34) [above = of 22, xshift = 0.55cm, yshift = -0.5cm] {$x_{i2}$};
		\node[state] (35) [above = of 23, xshift = -0.55cm, yshift = -0.5cm] {$\theta_3$};
		\node[state] (36) [above = of 23, xshift = 0.55cm, yshift = -0.5cm] {$x_{i3}$};

		\draw (31)--(21);
		\draw (32)--(21);
		\draw[color = julia2] (33)--(22);
		\draw (34)--(22);
		\draw (35)--(23);
		\draw (36)--(23);
		\draw (21)--(1);
		\draw[color = julia2] (22)--(1);
		\draw (23)--(1);
		\draw[color = julia2] (1)--(p);
		\draw[color = julia2] (p)--(l);
		\draw (Y)--(l);
	\end{tikzpicture}}
	\caption{
		Part of a GLM compute graph with three regression parameters.
		In the figure we compute $\ell_i$, the log-likelihood corresponding to data point $i$.
		Using caching techniques we reduce the complexity of Gibbs from $O(d^2)$ to $O(d)$.
		Our set of regression parameters after a coordinate update on $j=2$ with a Gibbs
		sampler is $\theta' = (\theta_1, \theta_2', \theta_3)$.
		To recompute $\ell_i$ after updating $\theta'_2$, we only need to 
		modify the entries highlighted in red. 
		Using caches updated by subtracting the old value and adding the new one at the `$+$' node,  this change can be performed in $O(1)$ operations, instead of
		$O(d)$ operations. Over $d$ coordinates, Gibbs then has a cost of $O(d)$ instead of
		$O(d^2)$.}
	\label{fig:CG_coordinate_update}
\end{figure}

Generalized linear models (GLMs) are among the most commonly used tools in contemporary 
Bayesian statistics \citep{gelman_bayesian_2013}. 
As a result, applied Bayesian statisticians require efficient 
algorithms to approximate the posterior distributions associated with GLM parameters, 
often relying on Markov chain Monte Carlo (MCMC).
In this paper we focus on two MCMC samplers: the Gibbs sampler and 
Hamiltonian Monte Carlo (HMC), popularized in statistics by \citet{geman_stochastic_1984} 
and \citet{neal_bayesian_1996}, respectively. 

Until around 2010, Gibbs sampling was the predominant option, as it formed the core of a ``first generation'' 
of probabilistic programming languages (PPLs) \citep{strumbelj_past_2024} such as BUGS \citep{lunn2009bugs} and JAGS \citep{plummer2003jags}.  
A large shift occurred in the early 2010s when HMC and 
its variant, the No-U-Turn Sampler (NUTS) \citep{hoffman_no-u-turn_2014}, were combined with reverse-mode 
automatic differentiation to power a ``second generation'' of 
PPLs such as Stan \citep{carpenter2017stan}. Subsequently, HMC largely replaced Gibbs as the 
focus of attention of the methodological, theoretical, and applied MCMC communities. 
A notable exception is Gibbs sampling under specific linear and conditionally 
Gaussian models \citep{papaspiliopoulos2020scalable,zanella2021multilevel}.

Our first contribution is the analysis of an algorithm for Gibbs sampling that speeds up inference of GLMs by a factor 
$O(d)$, where $d$ is the number of parameters, compared to the Gibbs implementation 
used by ``first generation'' PPLs. The main idea behind this speedup is to use 
information encoded in the \emph{compute graph} associated with the target log density---in contrast, 
previous Gibbs algorithms relied on  
graphical models \citep{jordan_graphical_2004}, which we show is a sub-optimal representation of GLMs for the purpose of efficient Gibbs sampling. 
The compute graph is a directed acyclic graph (DAG) encoding the dependencies between the operations involved in computing a 
function. See Figure~\ref{fig:CG_coordinate_update} for an example. Programming strategies 
to extract and manipulate such graphs are well developed, in large part because compute 
graphs also play a key role in reverse-mode automatic differentiation (see \citet{margossian_review_2019} 
for a review). 

This drastic $O(d)$ speedup prompted us to reappraise Gibbs sampling as a 
method of choice for approximating GLM posterior distributions. 
A holistic comparison of sampling algorithms requires taking into account not only the 
running time per iteration, but also the number of iterations to achieve one effective sample \citep{flegal_markov_2008}. 
To understand the latter, we use a combination of empirical results and theory. 
On the theoretical side, we consider normal models, where both HMC and Gibbs  
have well-developed results quantifying the number of iterations needed to achieve one effective sample 
\citep{roberts1997updating,ascolani2024entropy,Beskos2013OptimalTuningHMC,chen2023does}. 
See Section~\ref{sec:theory} for the rationale behind the choice of 
a normal model and an extended bibliography.
In the normal setting, the $O(d)$ runtime of our improved Gibbs sampler combined with past theory
yields a favourable $O(d)$ floating point operations (flops) per effective sample, 
compared with the higher $O(d^{5/4})$ flops per effective sample for HMC. 
While the $d^{1/4}$ advantage of Gibbs over HMC may seem minor, a key point is that Gibbs requires no adaptation to achieve this performance, 
while HMC needs to be finely tuned adaptively \citep{Livingstone2022barker}, implying that the practical rate for HMC may be even higher. 
Note also that previous, graphical model-based implementations of Gibbs sampling require
$O(d^2)$ flops per effective sample in the same setup. 

Examining scaling in $d$ alone does not provide a full story, even in the normal setting; 
the shape of the posterior contour lines also has a strong 
effect on the sampling performance of both Gibbs and HMC.
Various notions of \emph{condition number}  $\kappa$ are used 
\citep{langmore_condition_2019,hird2023precondition} 
to summarize the complexity brought by 
the shape of log-concave distributions. 
In gradient-based methods, a popular notion of condition number is the square of 
the broadest direction's scale to that of the most constrained.  For optimally-tuned 
HMC with non-constant integration time on Gaussian targets, 
the cost per effective sample is $O(d^{5/4} \kappa^{1/2})$ flops 
\citep{langmore_condition_2019,apers2022hamiltonian,wang2023accelerating,jiang23dissipation}. 
Understanding the condition number scaling of Gibbs is more nuanced, as
Gibbs sampling is invariant to axis-aligned stretching \citep{roman2014reparametrization},
but is sensitive to rotations (the situation is 
reversed for HMC \citep{neal2011hmc}). We develop a notion of 
\emph{residual condition number}, $\kappar$, to capture how Gibbs perceives the target's shape 
(see \cref{sec:preconditioning}). 
Crucially, $\kappar \le \kappa$. On the other hand, due to random walk behaviour, 
Gibbs' cost per effective sample scales as $O(d \kappar)$ flops, 
and in some cases, $\kappa = \kappar$, leading to a potentially worse or better scaling in the condition number for Gibbs compared to HMC, depending on the nature of the target. As a result, neither Gibbs sampling nor HMC dominate each other.

The relative importance of $\kappa$ and $d$ will in general be problem-dependent. 
To get some insight on how these quantities relate to each other in practice, we investigate sequences of 
posterior distributions obtained by 
subsampling an increasing number of covariates from real datasets. We find that various notions of  
condition number increase until $d \approx n$, after which they 
either stabilize or, surprisingly, can in certain cases decrease with $d$. This suggests that our efficient Gibbs sampling 
will be particularly effective in GLMs where $d \gg n$. 

We implemented our new Gibbs algorithm and benchmarked it using a collection of synthetic and real datasets. 
In \ndatasetswin\ out of the \ndatasetstotal\ datasets considered, our algorithm achieves a higher effective sample size (ESS) 
per second compared to Stan, with a speedup of up to a factor of 300.
\section{BACKGROUND}
\label{sec:background} 

In this section we introduce several key concepts for studying the Gibbs sampler and HMC 
when applied to GLMs. Throughout this paper, we denote the target distribution of interest on $\reals^d$
by $\pi$, with density $\pi(\theta) \propto e^{-U(\theta)}$, for $\theta \in \reals^d,$
with respect to Lebesgue measure on $\reals^d$. 
We assume that for all $\theta\in \reals^d$, $\nabla U(\theta)$ and 
$\nabla^2 U(\theta)$ are well-defined. 
The distribution $\pi$ is called \emph{strongly log-concave} if there exists a constant $a > 0$ such that 
for all $\theta \in \reals^d$, $\nabla^2 U(\theta) \succeq a I_d$, 
and \emph{strongly log-smooth} if there exists a constant $b > 0$ such that 
for all $\theta \in \reals^d$, $\|\nabla^2 U(\theta)\| \leq b < \infty$. 

\subsection{Gibbs sampling}
\label{sec:Gibbs}

In the literature, the term ``Gibbs sampler'' is associated with two key ideas: (1) moving a 
subset of coordinates while fixing the others; 
and (2), using the conditional distribution of the target distribution 
to perform such an update. 
The algorithm in Section~\ref{sec:fast_gibbs} is described for simplicity in the context of Gibbs samplers, 
but applies more generally to any ``\emph{Metropolis-within-Gibbs}'' \citep{chib1995understanding} sampler which are those based on (1) only. 
In the experiments, we use 
``slice sampling within Gibbs'', specifically, with doubling and shrinking \citep{neal_slice_2003}.
See \cref{sec:within-Gibbs} for more discussion on Gibbs versus ``within-Gibbs.'' 

For $\theta = (\theta_1, \ldots, \theta_d)$ and $j \in \cbra{1,2,\ldots,d}$, 
let $\theta_{-j}$ be the vector containing all components of $\theta$ except for $\theta_j$.
We define the conditional distributions $\pi_{j | -j}$, which correspond to the 
conditional distributions of $\theta_j$ given $\theta_{-j}$, where $\theta \sim \pi$. 
In Metropolis-within-Gibbs, each coordinate update is performed using a Markov kernel 
$K_{j|-j}$ that leaves $\pi_{j|-j}$ invariant. 
The \emph{Gibbs sampler} uses the Markov kernels $K_{j | -j}(x, \cdot) = \pi_{j|-j}(\cdot)$,
but more general kernels are commonly used as well \citep{neal_slice_2003}.
It remains to decide the order in which to update coordinates.
One popular approach is to use a deterministic update Gibbs sampler (DUGS) 
\citep{roberts1997updating,greenwood1998information}, where coordinates are updated in a fixed order.
For example, when $d=3$, the DUGS kernel is:
$\theta_1' \sim K_{1|-1}(\cdot | \theta_2, \theta_3),$ followed by 
 $\theta_2' \sim K_{2|-2}(\cdot | \theta_1', \theta_3)$, and finally, 
 $\theta_3' \sim K_{3|-3}(\cdot | \theta_1', \theta_2')$.
We write $\DUGS$ for the deterministic alternation of the $d$ Gibbs kernels $K_{j | -j}(x, \cdot) = \pi_{j|-j}(\cdot)$. 
Options beyond DUGS include the \emph{random scan} Gibbs sampler, which chooses coordinates to update randomly,
and \emph{block} Gibbs samplers, which chooses several coordinates to update at once and can improve convergence \citep{li2005convergence}.
In this work, we focus primarily on $\DUGS$,
as it yields a smaller asymptotic variance
than random scan Gibbs in some problem classes 
\citep{greenwood1998information,qin2022convergence,andrieu2016random}---although a 
definitive conclusion is more nuanced \citep{roberts2015surprising,he2016scan}. 
Note also that random scan Gibbs would yield similar scaling results for the targets 
we consider \citep{ascolani2024entropy} (see also \cref{app:sweep-type} for an empirical 
demonstration).

\subsection{Hamiltonian Monte Carlo}
\label{sec:HMC}

Hamiltonian Monte Carlo (HMC) is an MCMC algorithm that makes use of gradient information 
of the target density and introduces a momentum in the sampling process 
to provide efficient exploration of the state space \citep{neal2011hmc}. 
Calculating the trajectory of HMC samples requires solving a differential equation, 
which is approximated using numerical integrators in practice.
We introduce HMC in two phases: using idealized trajectories, which assume that 
exact numerical integrators exists; and with the leapfrog integrator, a popular 
numerical integrator in the case of HMC. 
In both cases, the HMC kernel has an invariant distribution on an augmented space
with momentum variables $p \in \reals^d$, with joint density $\bar\pi(\theta, p) = \pi(\theta) \cdot \Norm(p \mid 0, M),$
which admits $\pi$ as the $\theta$-marginal. Here, $M$ is the covariance matrix of the Gaussian momentum, 
which is often referred to as the \emph{mass matrix} of HMC. 

For \emph{idealized HMC}, we initialize with a starting point 
$\theta^{(0)} = \theta(0)$ at time $t=0$ and some momentum $p(0) \sim \Norm(0, M)$. 
We specify a (possibly random) integration time $\tau$ and solve for $\cbra{\theta(t)}_{t \in [0,\tau]}$ from 
\[
  \label{eq:ideal_HMC_trajectory}
  \theta'(t) = M^{-1}\cdot p(t), \qquad p'(t) = -\nabla U(\theta(t)),
\]
which corresponds to Hamiltonian dynamics.
We then record the draw $\theta^{(1)} \gets \theta(\tau)$, 
reset $\theta(0) \gets \theta(\tau)$, and draw a new 
$p(0) \sim \Norm(0, M)$. 
Repeating the process, we obtain a sequence of draws $\theta^{(0)}, \theta^{(1)}, \theta^{(2)}, \ldots$. 

It is not generally possible to simulate the trajectory in \cref{eq:ideal_HMC_trajectory} 
exactly, so numerical integration is used. For a given starting point 
$(\theta,p)$, integration step size $\eps > 0$, and number of steps $s$,
the leapfrog integrator $L_\eps$ \citep{neal2011hmc} is applied $s$ times,
where $(\theta', p') = L_\eps(\theta, p)$ is defined by
$p'_{\half} = p - \frac{\eps}{2} \nabla U(\theta),$ followed by 
$\theta'  = \theta + \eps M^{-1} \cdot p'_{\half},$, and finally, 
$p' =  p'_{\half} - \frac{\eps}{2} \nabla U(\theta').$
Applying $L_\eps^s(\theta,p)$ ($L_\eps$ applied $s$ times to $(\theta, p)$) 
we obtain a proposed point $(\sttheta, \stp)$ that approximates the idealized HMC point $(\theta(\tau), p(\tau))$ 
at time $\tau = \eps s$.
A Metropolis--Hastings (MH) accept-reject step is then introduced in order to leave 
$\bar\pi$ invariant; the proposal $(\sttheta, \stp)$ is accepted with the standard Metropolis probability denoted $\alpha((\theta,p), (\sttheta, \stp))$ and otherwise we remain at $(\theta,p)$.

\subsection{Generalized linear models} 
\label{sec:GLMs} 

Suppose we are given $n$ observed pairs of covariates $x_i \in \reals^d$ and responses $y_i\in \scY\subset \reals$, 
$i \in \{1, 2, \dots, n\}$. 
The independent random responses $Y_i$ are assumed to have a 
conditional density $f_{Y|X}$ parameterized in terms of the mean of the distribution; 
in a generalized linear model (GLM), the mean of the distribution of response $Y_i$ is assumed to depend 
only on the \emph{linear predictor} $x_i^\top \theta$, where $\theta \in \reals^d$.
We fix some \emph{inverse link function}, $\mu$, and model the mean $\E[Y_i] = \mu(x_i^\top \theta) \equiv \mu_i.$
The full likelihood is then $\scL(\theta) = \prod_{i=1}^n f_{Y|X}(y_i \mid \mu_i),$
and the log-likelihood corresponding to data point $(x_i, y_i)$ is $\ell_i = \log f_{Y|X}(y_i \mid \mu_i)$.
In the Bayesian framework, a prior with density $\pi_0$ is specified on the regression parameters,
such that one obtains a posterior distribution with  density $\pi(\theta) \propto \pi_0(\theta) \cdot \scL(\theta).$

\subsection{Condition numbers and preconditioning} 
\label{sec:preconditioning} 

The \textit{condition number} of a given positive definite matrix $M$ is defined as 
$\kappa(M):=\lambda_\text{max}(M)/\lambda_\text{min}(M)$ where $\lambda_\text{max}(M)$ 
and $\lambda_\text{min}(M)$ are the largest and smallest eigenvalues of $M$, respectively.
We can generalize this definition to accommodate strongly log-concave distributions. 
For a strongly log-concave distribution $\pi$,
the condition number of $\pi$ can be defined as:
\[ \label{eq:condnumberdef}
    \kappa(\pi):=\sup _{\theta \in \mathbb{R}^d}\left\|\nabla^2 U(\theta)\right\| \sup _{\theta \in \mathbb{R}^d}\left\|\nabla^2 U(\theta)^{-1}\right\|.
\]
In particular, when $\pi$ is Gaussian $\Norm(\mu, \Sigma)$, we have 
$\kappa(\pi) = \kappa(\Sigma) = \lambda_\text{max}(\Sigma)/\lambda_\text{min}(\Sigma)$.
Moving forward, we refer to $\kappa(\pi)$ as the \textit{raw condition number};
we simply denote this by $\kappa$ when there is no ambiguity.

It is common to transform the target distribution $\pi$ with a preconditioning 
matrix $A$ in order to try to reduce the condition number. 
That is, given $\theta \sim \pi$ and a full rank $d \times d$ matrix $A$, 
we define $\pi_A$ such that $\theta_A = A \theta \sim \pi_A$. In the context of HMC,
preconditioning with $A$ is equivalent to setting the mass matrix to $A^\top A$.
One common approach to preconditioning is to set $A = \diag \rbra{\Var_\pi[\theta_j]^{-1/2}}.$
In this case, we refer to $\kappa(\pi_A)$ as the \textit{correlation condition number}, 
denoted as $\kappac(\pi)$,
since the covariance and correlation matrix of $\pi_A$ are the same. 
It is often implicitly assumed that $\kappac\le\kappa$, but
this is not always true \citep{hird2023precondition}, 
as we also see in our experiments.

We finally introduce the notion of the \emph{residual condition number}, $\kappar(\pi)$,
which is defined as $\kappar(\pi):=\inf_{A \in \mathcal{D}}\kappa(\pi_A),$
where $\mathcal{D}$ is the set of full rank $d\times d$ diagonal matrices. 
By definition, $\kappar\le\kappa$. We will demonstrate in \cref{sec:theory} that the convergence rate 
for Gibbs sampling depends on $\kappar$, as opposed to $\kappa$.

\section{COMPUTE GRAPH GIBBS} 
\label{sec:fast_gibbs} 

In this section, we present a fast algorithm for Gibbs sampling exploiting 
the structure of the compute graph. 
When this algorithm is applied to Gibbs sampling of GLMs, 
it reduces the computation time for a full scan of $d$ updates from $O(d^2 n)$ to $O(d n)$. 
Special cases of this algorithm have been applied to Gibbs 
sampling of
specific models (see, e.g., \citet{mahani2015simd}, Equations 14--16), but not to probabilistic programming languages as far 
as we are aware. Our compute graph formulation makes it possible to 
automate and generalize the use of the algorithm presented in this section. 

To illustrate our technique, we turn to logistic regression with parameters 
$\theta = (\theta_1, \ldots, \theta_d)$ and data points $\cbra{(x_i,y_i)}_{i=1}^n$. 
The likelihood for this model is given by 
\[
  \label{eq:logistic_likelihood}
  \scL(\theta) &= \prod_{i=1}^n p_i(\theta)^{y_i} (1-p_i(\theta))^{1-y_i}, \\
  p_i(\theta)  &= \frac{1}{1+\exp\rbra*{- x_{i}^\top\theta}}.
\]
We now introduce a flavour of \emph{compute graph} suitable for our purpose, 
namely a type of directed graph associated with a 
function $f(\theta_1, \dots, \theta_d)$---see \cref{fig:CG_coordinate_update} for an example in the GLM context. 
First, for each variable $\theta_j$, 
we assume there is a corresponding \emph{input} node in the graph such that no edge points into it. 
We call the graph vertices that are not input nodes the \emph{compute nodes}. 
For each compute node $n$, we assume there is an associated operation, i.e., a function 
that takes as input the values given by the nodes $n'$ such that $(n' \to n)$ is an edge 
in the graph. We assume the compute graph has a single \emph{sink node}, i.e., a node which has 
no outgoing edge. We also posit that the composition of the operations along the compute graph from inputs to 
sink yields a function that is identical to $f(\theta_1, \dots, \theta_d)$. 

In \cref{fig:CG_coordinate_update}, we see that the part of the compute graph for calculating the likelihood 
associated with data point $x_i$ revolves around computing the term $x_i^\top \theta$.  
After a single-coordinate update of $\theta$, as would be done with the Gibbs sampler,
many of the compute graph nodes hold the same value before and after the update (in \cref{fig:CG_coordinate_update}, those that change are highlighted in red). 
We explain next how these invariant values allow us to attain an $O(d)$ speedup of the Gibbs sampler for GLMs.

We now introduce our approach for computing the likelihood of GLMs, 
which we call compute graph Gibbs (CGGibbs). This approach
is based on the following idea: cache the scalar values of the linear predictors 
$x_i^\top \theta$ for each data point $i$ in a vector of length $n$, $\texttt{cache}$. 
This requires only $O(n)$ memory (this is negligible since the design matrix already requires storing $d \cdot n$ entries), 
but yields an $O(d)$ computational speedup. This approach is possible because 
the Gibbs sampler considers coordinate-wise updates of the form 
$\theta = (\theta_1, \ldots, \theta_j, \ldots, \theta_d) \mapsto 
\theta' = (\theta_1, \ldots, \theta'_j, \ldots, \theta_d)$, 
and consequently likelihood updates have a special structure.
For such updates, we have the following convenient decomposition 
\[
  x_i^\top \theta' 
  &= \underbrace{\sum_{k=1}^d \theta_k x_{ik}}_{\texttt{cache}[i]} - \theta_j x_{ij} + \theta_j' x_{ij}.
\]
This new linear predictor for a given data point can be computed in $O(1)$ time with caching, compared to 
$O(d)$ time without caching.
From here, one evaluation of the full likelihood comes at an $O(n)$ cost, 
and a full sweep over $d$ coordinates
is thus an $O(dn)$ cost, instead of the usual $O(d^2 n)$ that would be incurred without caching.
The resulting speedup in terms of the number of parameters, $d$, 
allows us to apply this new approach to Bayesian inference with Gibbs sampling 
to high-dimensional regression problems, including ones where $d \gg n$. 

Although this simple optimization is obvious from the \emph{compute graph}, presented in 
\cref{fig:CG_coordinate_update}, it is not obvious from the \emph{directed graphical model} corresponding 
to the given regression problem in \cref{fig:DAG_regression}. 
Probabilistic programming languages (PPLs) incorporating Gibbs samplers often 
perform optimizations in evaluating ratios of likelihoods for Markov kernel proposals 
with respect to graphical models, such as the one in \cref{fig:DAG_regression} \citep{lunn2009bugs}. 
However, these graphs only reveal the dependence structure of random variables in the model 
and decompositions of the likelihood, but they do not yield insight into fine-grained 
optimization of computations such as in \cref{fig:CG_coordinate_update}. 
In our case, this ``fine-grained'' optimization yields a substantial $O(d)$ improvement in 
computation time with the introduction of simple caching techniques.

\begin{figure}[t]
  \centering 
  \scalebox{0.7}{
  \begin{tikzpicture}
    \node[state,fill=lightgray] (1) {$Y_1$};
    \node[state,fill=lightgray] (2) [right = of 1] {$Y_2$};
    \node[state,fill=lightgray] (3) [right = of 2] {$Y_3$};
    \node[state,fill=lightgray] (4) [right = of 3] {$Y_4$};
    \node[state,fill=lightgray] (5) [right = of 4] {$Y_5$};
    \node[state] (b1) [above = of 2, yshift = 5.0] {$\theta_1$};
    \node[state] (b2) [right = of b1] {$\theta_2$};
    \node[state] (b3) [right = of b2] {$\theta_3$};

    \path (b1) edge node[above] {} (3);
    \path (b2) edge node[above] {} (3);
    \path (b3) edge node[above] {} (3);
    \draw[opacity = 0.3] (b1)--(1);
    \draw[opacity = 0.3] (b1)--(2);
    \draw[opacity = 0.3] (b1)--(3);
    \draw[opacity = 0.3] (b1)--(4);
    \draw[opacity = 0.3] (b1)--(5);

    \draw[opacity = 0.3] (b2)--(1);
    \draw[opacity = 0.3] (b2)--(2);
    \draw[opacity = 0.3] (b2)--(3);
    \draw[opacity = 0.3] (b2)--(4);
    \draw[opacity = 0.3] (b2)--(5);

    \draw[opacity = 0.3] (b3)--(1);
    \draw[opacity = 0.3] (b3)--(2);
    \draw[opacity = 0.3] (b3)--(3);
    \draw[opacity = 0.3] (b3)--(4);
    \draw[opacity = 0.3] (b3)--(5);
 \end{tikzpicture}}
  \caption{The \emph{directed probabilistic graphical model} 
  for the same regression problem as \cref{fig:CG_coordinate_update} with five data points $(x_i, Y_i)$ 
  and three regression parameters $\theta_j$. 
  Only edges into $Y_3$ are emphasized and the covariates $x_i$ are assumed fixed. 
  Observed variables are shaded in gray.
  Each time a $\theta_j$ is changed, we have to recompute the likelihood from scratch 
  for this representation as opposed to updating the cache at the `$+$' node found in 
  the \emph{compute graph} representation in \cref{fig:CG_coordinate_update}.}
  \label{fig:DAG_regression}
\end{figure}

\section{THEORY} 
\label{sec:theory} 

Since GLMs come in many flavours of priors and likelihoods, we would ideally perform theoretical 
comparison of Gibbs and HMC  
under broad assumptions such as log-concavity. 
Unfortunately, current theoretical results for the log-concave setting 
are not mature enough---we seek to avoid the fallacy of comparing algorithms 
via loose complexity upper bounds.
Specifically, current results for HMC and Gibbs on log-concave targets fall 
short on various aspects, such as failing to take into account both condition number 
and dimension simultaneously  \citep{wang2017convergence,wang2019convergence,wang2014convergence,wang2020convergence} 
or focusing on fixed integration time HMC instead of the non-constant integration time 
variants used in practice \citep{chen2019optimal}. 
In fact, the scaling of randomized or 
dynamic HMC as a function of the condition number for log-concave targets is still an open problem 
as of the time of writing, with different authors positing conflicting conjectures \citep{lee2020logsmooth,apers2022hamiltonian}.

Instead, we focus on normal models in this section. On one hand, Bayesian GLM posterior distributions 
are not exactly normal, but on the other hand,
based on Bernstein-von Mises theorems \citep{vaart_asymptotic_1998}, many are well approximated by normal 
distributions. Other Bayesian GLM posterior 
distributions are not 
approximately normal, so we complement the results in this section with experiments 
on Bayesian GLMs with highly non-normal posterior distributions (\cref{sec:HSP_example}).

\subsection{Convergence rates and scaling} 
\label{sec:scaling}

\paragraph{Gibbs convergence rate} Building on \citet{roberts1997updating}, 
we first establish the convergence rate of the idealized DUGS algorithm on
Gaussian distributions with respect to various divergences: 
1-Wasserstein, 2-Wasserstein distance, and the Pearson-$\chi^2$ divergence, 
denoted $\TV, \mathrm{W}_1, \mathrm{W_2}$ and $\chi^2$, respectively 
(see \cref{app:theory-practice} for background on these divergences and how they relate 
to the empirical results in \cref{sec:experiments}).

\bthm 
\label{thm:Gibbs_convergence}
Let $\pi = \Norm(\mu, \Sigma)$ with precision matrix $\Sigma^{-1}$ having only 
non-positive off-diagonal elements. For any initial point $x \in \reals^d$ and any 
$\mathrm{D} \in \{\TV, \mathrm{W}_1, \mathrm{W_2}, \chi^2\}$, we have
\[
  \mathrm{D}(\DUGS^t(x, \cdot), \pi) 
  = O\left(\exp\left(-\frac{t}{\kappa(\Sigma)}\right)\right), \quad t \to \infty.
\]
\ethm
\cref{thm:Gibbs_convergence} shows that the contraction rate of DUGS is independent of $d$ 
for all four divergences. Concurrently to our work, \citet{ascolani2024entropy} has proved
a Kullback-Leibler divergence bound with the same convergence rate as in \cref{thm:Gibbs_convergence} for
strongly log-concave and log-smooth targets and random sweep Gibbs samplers.
Currently available bounds for randomized HMC do not appear as tight
for strongly log-concave and log-smooth targets (see discussion on HMC convergence rate in
the next paragraphs). Therefore, the comparison of theoretical results for Gibbs and HMC
in the Gaussian case is still informative.

\paragraph{Ideal HMC convergence rate} 
Most scaling results for \emph{Metropolized HMC} are presented in
terms of \emph{mixing times} (which we review in the next section), and not on \emph{convergence 
rates}. See \cref{sec:rate_vs_mix_time} for more discussion on convergence rates versus mixing times. 
There are, however, several studies on the convergence rates of \emph{idealized 
HMC}. For example, \citet{chen2019optimal} provided a tight 
convergence rate bound of
$O\left(1 - \frac{1}{16\kappa(\pi)}\right)^t$ 
for idealized HMC with constant integration time on strongly log-concave and log-smooth targets $\pi$.
However, \citet{wang2023accelerating} achieved a $\mathrm{W}_2$ contraction rate
with improved $\kappa$
scaling---$O\left(1-\Theta\left(\frac{1}{\sqrt{\kappa}}\right)\right)^t$---for
idealized HMC on Gaussian targets using a time-varying integration time called
\emph{Chebyshev integration time}.
\citet{jiang23dissipation} obtained the same accelerated rate on Gaussian targets 
using a random integration time combined with partial momentum refreshment.
Whether these $O\left(1-\Theta\left(\frac{1}{\sqrt{\kappa}}\right)\right)^t$ rates of idealized HMC 
can be generalized to non-Gaussian targets remains an open question.
Furthermore, although the rates in \citet{wang2023accelerating,jiang23dissipation} are dimension-independent, 
idealized HMC is not implementable in practice since it requires exact ODE simulation. 

\paragraph{Metropolized HMC mixing time}
The current state-of-the-art mixing rate for Metropolized HMC on general log-concave 
and log-smooth targets is $O(\kappa d^{1/4} \log(1/\eps))$ gradient queries
for an $\eps$-level error in total variation distance \citep{chen2023does}.
The $O(d^{1/4})$ scaling (without known dependence on $\kappa$) 
was originally established in \citet{Beskos2013OptimalTuningHMC} 
for separable log-concave, log-smooth targets. 

In another development, \citet{chen2020fast} proved a mixing time of
$O(\kappa d^{11/12})$ for general log-concave and log-smooth targets.
Additionally, \citet{lee2020logsmooth} showed
that HMC with a single leapfrog step per iteration has a mixing time of
$O(\kappa d)$, and they conjectured that the $O(\kappa)$ dependence might be
tight in this setting.

This dependence on the condition number has been improved in the case
of Gaussian targets. Specifically, \citet{apers2022hamiltonian} achieved a
mixing time of $O(\kappa^{1/2} d^{1/4})$ by employing randomized integration
steps at each iteration. This result mirrors the improved condition number
scaling in the idealized HMC rates, and the authors of this study
conjecture that this enhanced $\kappa$ scaling could potentially generalize to
broader classes of log-concave and log-smooth targets, although this is yet to 
be studied.

\subsection{Influence of diagonal preconditioning}\label{sec:tuning}

The scaling results presented in \cref{sec:scaling} suggest that Gibbs sampling
has better dimension scaling than Metropolized HMC with non-constant trajectory lengths, 
at least in the Gaussian case. In terms of condition number scaling 
the situation is more nuanced, and a full characterization requires a careful 
analysis of diagonal preconditioning, which we now discuss.

A first observation is that focusing solely on $\kappa(\pi)$, the 
condition number of the untransformed target, might not adequately represent 
the practical performance of the two samplers. 
While the observation described in the previous sentence is true for both samplers, its underlying cause  is quite distinct for Gibbs and HMC, 
and so is the appropriate notion of condition number in 
each case.  

For HMC, the standard practice is to fit a diagonal preconditioning matrix via adaptive MCMC;
restricting to diagonal matrices ensures that the compute cost per leapfrog step stays linear in $d$. 
For instance, NUTS as implemented 
in Stan by default adopts diagonal preconditioning using estimated marginal standard deviations. 
Hence, for HMC run for enough iterations, the scaling in condition number 
will depend on the correlation condition number $\kappac$
introduced in \cref{sec:preconditioning}, but with the 
caveat that marginal estimates need to be learned. Therefore, for 
early iterations the dependence will be on $\kappa$ rather than $\kappac$. 
On the other hand, thanks to the suppression of random walks brought by HMC with 
long trajectories, it is possible to achieve $\kappa^{1/2}$ scaling as discussed in 
\cref{sec:scaling}, 
provided that HMC's key tuning parameters, the integrator step size and 
trajectory length, are well-tuned. 

The situation for Gibbs brings a mix of good news and bad news,
as we formalize in  \cref{prop:conditioning_invariance_Gibbs}. 
On the negative side, the dependence grows linearly 
in the condition number instead of as a square root. 
On the positive side, because Gibbs is invariant to axis-aligned 
stretching, it is as if Gibbs automatically uses the optimal diagonal preconditioner, 
without having to explicitly learn it. 

\bprop
\label{prop:conditioning_invariance_Gibbs}
Under the same conditions as \cref{thm:Gibbs_convergence}, 
for any initial point $x \in \reals^d$ and any 
$\mathrm{D} \in \{\TV, \mathrm{W}_1, \mathrm{W_2}, \chi^2\}$,
\[
    \mathrm{D}(\DUGS^t(x, \cdot), \pi) 
    = O\left(\exp\left(-\frac{t}{\kappar(\Sigma)}\right)\right), \quad t \to \infty.
\]
\eprop

In contrast, \emph{HMC may even experience negative effects from diagonal
preconditioning} (or equivalently, mass matrix adaptation). 
This was pointed out previously by \citet[Sec 3.5.3]{hird2023precondition}, who provide an 
instance where diagonal preconditioning using the target standard deviation 
results in a worse condition number, even in the Gaussian case.

Even when such preconditioning theoretically offers benefits, current mass matrix
adaptation methods rely on moment-based estimates, which can take a considerable
amount of time to become accurate. Therefore, one might not achieve the
idealized $\kappa$ rate in practice with HMC due to suboptimal mass matrix adaptation 
(see \cref{sec:cond_num_speed}).

Other works have also considered condition numbers resulting from 
different preconditioning strategies.
\citet{ascolani2024entropy} used the condition number after each dimension
has been scaled by the square root of their respective smoothness constant.
\citet{hird2023precondition} showed how the condition number is affected by
various linear preconditioning strategies such as scaling by the square root of
the Fisher information matrix, the covariance matrix or using the R matrix from
the QR decomposition, etc.

\section{EXPERIMENTS} 
\label{sec:experiments}

In this section, we conduct experiments showing the performance gain of CGGibbs over other 
popular Gibbs sampler implementations as well as comparing the efficiency of CGGibbs and NUTS 
on real datasets. The source code for these experiments can be found at 
\url{https://github.com/UBC-Stat-ML/gibbs-vs-hmc-mev}. All experiments were conducted on 
the ARC Sockeye computer cluster at the University of British Columbia.
Additional experiments, details of the experimental setup and instructions for reproducibility are available in 
\cref{app:exp}.

\subsection{Compute graph Gibbs is faster than prevailing Gibbs implementations}
\label{sec:synthetic_experiments}

\begin{figure}[t]
  \centering
  \includegraphics[width=0.38\textwidth]{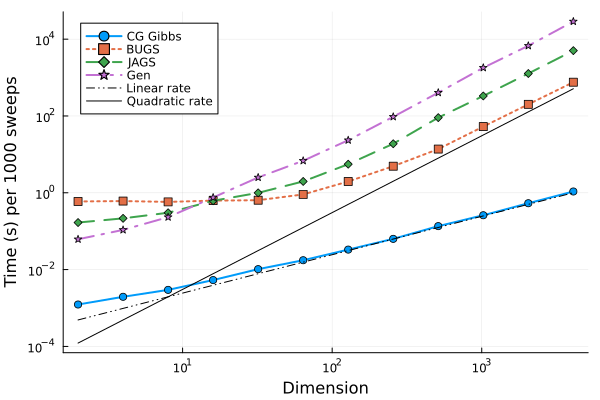}
  \caption{Wall-clock time (in seconds) taken to perform 1000 sweeps versus dimension for various Gibbs sampler implementations 
  on synthetic logistic regression datasets of increasing dimensionality (log-log scale, lower is better). }
  \label{fig:Gibbs_scaling}
\end{figure}

We first study the running time of various implementations of within-Gibbs samplers: our CGGibbs sampler, 
as well as the popular MultiBUGS v2.0 \citep{lunn2009bugs}, JAGS v4.3.2 \citep{plummer2003jags} and Gen.jl v0.4.7 
\citep{cusumano2019gen} software packages. The goal is to confirm that previous implementations 
of within-Gibbs samplers scale in $O(d^2)$ per sweep, versus $O(d)$ for our method. For this experiment, we consider a sequence of synthetic
logistic regression datasets with increasing dimension 
$d \in \cbra{2^1, \ldots, 2^{12}}$. Each parameter has a Gaussian prior with standard deviation 10. Since in this first experiment we restrict ourselves to Gibbs samplers, we use the time taken 
to run 1000 sweeps as a representation of the computational complexity. 
The results, presented in \cref{fig:Gibbs_scaling}, 
show that CGGibbs does indeed achieve an $O(d)$ scaling, 
while other currently available and commonly used Gibbs samplers have an 
undesirable $O(d^2)$ scaling. 

\subsection{Empirical scaling of compute graph Gibbs and NUTS}\label{sec:empirical-scaling}

\begin{figure}[t]
  \centering
  \includegraphics[width=0.38\textwidth]{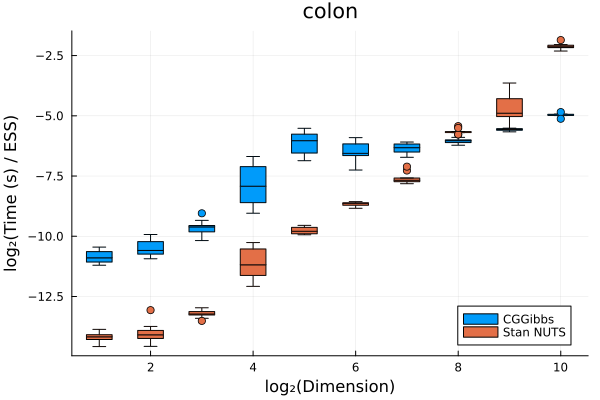}
  \caption{Wall-clock time (in seconds) per ESS for CGGibbs
    and Stan NUTS as a function of the number of subsampled predictors (log-log scale, lower is better) from a colon cancer gene expression dataset \citep{alon1999broad}. Box plots summarize 10 replicates. }
  \label{fig:dim_scale_colon}
\end{figure}

Next, we compare the wall-clock time per ESS as a function of 
the number of covariates for the Gibbs and Stan NUTS (2.35.0) \citep{carpenter2017stan} samplers. We exclude adaptation time from timings and focus on the 
stationary regime for simplicity. 
In the main text, we summarize the ESS performance by taking the median across all test functions $\theta \mapsto \theta_i$ (first moment of each marginal) and $\theta \mapsto \theta^2_i$ (squared marginal moments), see also \cref{app:exp} for 
results where ESS is summarized with the minimum ESS across test functions.
We show results on the colon cancer dataset \citep{alon1999broad} in the main 
text, see \cref{sec:more_dim_scale} for similar experiments on other datasets. 
For each dataset, we create a sequence of 
logistic regression problems with an increasing number of predictors and obtain samples from the 
posterior of each of these problems. Specifically, 
we shuffle uniformly at random the order of the covariates and choose an increasing prefix of $d \in \cbra{2^1,2^2, \ldots, 2^{10},2000}$ features from 
the permuted dataset (the colon dataset has 2000 features) to be predictors for the response and use 
an isotropic normal prior with standard deviation 10 for the parameters. 
Note that by adding covariates we change not only the dimensionality but also the 
condition number of the target distribution. 
The results are shown in \cref{fig:dim_scale_colon}, where we see that the 
efficiency of CGGibbs scales favourably as a function of the number of covariates compared to Stan NUTS. 

\begin{figure}[t]
  \centering
  \includegraphics[width=0.38\textwidth]{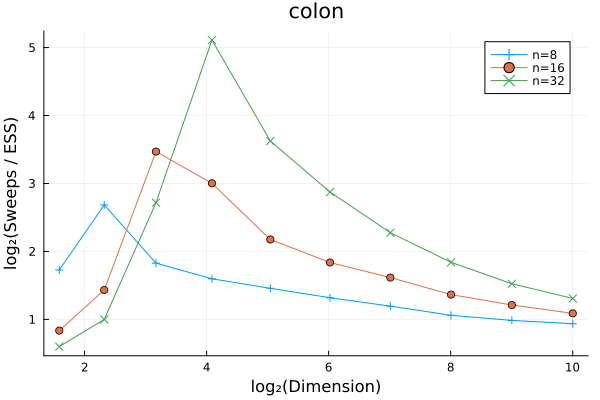}
  \caption{Dimensional scaling of the number of Gibbs sweeps to achieve one ESS 
  for different subsampled size of data points (series colours) and predictors (abscissa). }
  \label{fig:dim_vs_size_scaling}
\end{figure}

Upon closer inspection of \cref{fig:dim_scale_colon}, there appears to be a change in behaviour 
in the performance of CGGibbs when $d \approx n$ (the colon dataset has $n=62$ observations). 
To investigate this, we perform the same experiment but with a varying 
number of observations with $n \in \cbra{2^3, 2^4, 2^5}$ and record the number of sweeps 
per median ESS. \cref{fig:dim_vs_size_scaling} confirms that there is 
again a similar transition when $d<n$ to $d>n$. This behaviour is partially explained 
with the stabilization (or decrease) of various notions of condition number when 
$d$ increases past $n$ (see \cref{sec:cond_num_scale}). Interestingly, the decrease in 
sweeps per ESS is similar to a result from \citet[Theorem 2]{qin2024spectral}, 
where they developed a lower bound for the convergence rate of a random scan Gibbs sampler 
for submodels that depend on the full model. We further investigate this behaviour using 
controlled experiments with synthetic data in \cref{sec:irrelevant_params}.

\subsection{Panel of datasets} 
\label{sec:real_data_experiments} 

We compare the time per ESS 
for CGGibbs and Stan NUTS on \ndatasetstotal\ binary classification datasets. These datasets include 
newsgroup datasets \citep{lang1995newsweeder},
gene expression datasets 
\citep{golub1999molecular,alon1999broad,singh2002gene,spira2007airway,freije2004gene},
a mass spectrometry data set \citep{guyon2004result}
a digit recognition data set \citep{guyon2004result}
and an artificial dataset \citep{guyon2004result}.
We use a logistic regression
model on each dataset with two types of priors: 
a Gaussian prior with a standard deviation of 10 
for each of the parameters and a horseshoe prior \citep{carvalho2009horseshoe}. 
Inference for each combination of dataset and prior is repeated 30 times with different seeds. 
To avoid unreliable ESS estimates, we only report ESS estimates when the trace 
is sufficiently long to achieve an ESS of at least 100. 
More details on the data and model can be found in \cref{sec:real_data_details}.

\begin{figure}[t]
  \centering
  \includegraphics[width=0.35\textwidth]{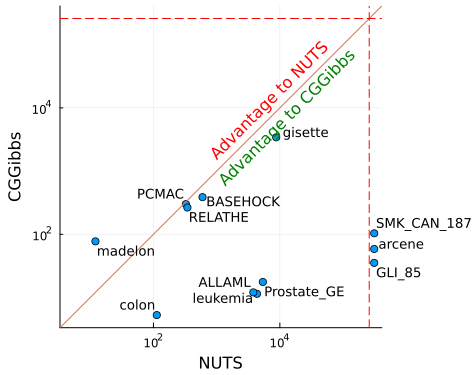}
  \caption{Median (across 30 replicates) wall-clock time (in seconds) per 100 ESS for CGGibbs and Stan NUTS 
  on \ndatasetstotal\ logistic regression real datasets with a Gaussian prior. 
  Each point is a dataset: its x-axis coordinate denotes the median time for NUTS, its y-axis 
  coordinate, for CGGibbs. 
   The region beyond 
  the dashed lines indicates that the corresponding sampler has not reached a minimum of 100 
  ESS within three days.}
  \label{fig:real_data_ess_gaussian}
\end{figure}

\begin{figure}[t]
  \centering
  \includegraphics[width=0.35\textwidth]{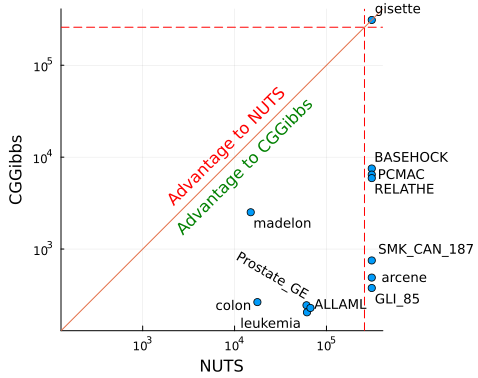}
  \caption{Similar plot to \cref{fig:real_data_ess_gaussian} but for the horseshoe prior.}
  \label{fig:real_data_ess_HSP}
\end{figure}

\begin{figure}[t]
  \centering
  \includegraphics[width=0.4\textwidth]{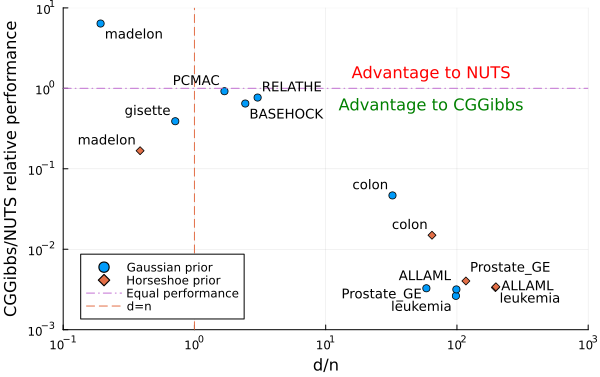}
  \caption{Ratio of median time (in seconds) per ESS between CGGibbs and Stan NUTS
  versus the $d/n$ ratio. Each point represents a combination of dataset and prior where both samplers 
  reached 100 minimum ESS within 3 days. 
  The dashed line indicates where $d=n$ and the dash dotted line indicates the threshold 
  of equal performance between the two samplers.}
  \label{fig:d/n}
\end{figure}

The outcome of these experiments is summarized in
\cref{fig:real_data_ess_gaussian,fig:real_data_ess_HSP}. Overall, CGGibbs outperforms 
Stan NUTS in almost all datasets and is more than 300 times faster than Stan NUTS in time 
per ESS for the best case. Furthermore, 
\cref{fig:d/n} suggests a correlation
between the $d/n$ ratio and the performance of CGGibbs relative to NUTS.

\section{DISCUSSION} 
\label{sec:discussion}

Both our CGGibbs algorithm and reverse-mode automatic differentiation, 
(the key ingredient to automate the use of HMC in PPLs \citep{margossian_review_2019}) are 
based on related but distinct notions of compute graphs. 
While we have focused on GLMs for concreteness, the efficient update of the 
compute graph described in \cref{sec:fast_gibbs} appears to apply more generally. 
Automatic processing of arbitrary compute graphs for efficient CGGibbs updates 
is potentially simpler than the machinery needed for reverse-mode automatic differentiation: 
in the former case, only certain reduce operations such as addition need special treatment, 
whereas in the latter case, all primitive operations need to be handled 
individually (i.e., each provided with an adjoint).  
Beyond GLMs, other examples of uses of CGGibbs include models with 
sufficient statistics and models of infinite dimensionality. 
Another interesting future direction is the analysis of algorithms combining the 
strengths of both Gibbs and HMC, such as optimal designs of 
block HMC-within-Gibbs algorithms.

\subsubsection*{Acknowledgements}
ABC and TC acknowledge the support of an NSERC Discovery Grant.
NS acknowledges the support of a Vanier Canada Graduate Scholarship. 
We additionally acknowledge use of the ARC Sockeye computing platform from the
University of British Columbia.

\bibliographystyle{apalike}
\bibliography{main.bib}

\section*{Checklist}

 \begin{enumerate}

 \item For all models and algorithms presented, check if you include:
 \begin{enumerate}
   \item A clear description of the mathematical setting, assumptions, algorithm, and/or model. [\textbf{Yes}/No/Not Applicable]
   \item An analysis of the properties and complexity (time, space, sample size) of any algorithm. [\textbf{Yes}/No/Not Applicable]
   \item (Optional) Anonymized source code, with specification of all dependencies, including external libraries. [\textbf{Yes}/No/Not Applicable]
 \end{enumerate}

 \item For any theoretical claim, check if you include:
 \begin{enumerate}
   \item Statements of the full set of assumptions of all theoretical results. [\textbf{Yes}/No/Not Applicable]
   \item Complete proofs of all theoretical results. [\textbf{Yes}/No/Not Applicable]
   \item Clear explanations of any assumptions. [\textbf{Yes}/No/Not Applicable]     
 \end{enumerate}

 \item For all figures and tables that present empirical results, check if you include:
 \begin{enumerate}
   \item The code, data, and instructions needed to reproduce the main experimental results (either in the supplemental material or as a URL). [\textbf{Yes}/No/Not Applicable]
   \item All the training details (e.g., data splits, hyperparameters, how they were chosen). [\textbf{Yes}/No/Not Applicable]
         \item A clear definition of the specific measure or statistics and error bars (e.g., with respect to the random seed after running experiments multiple times). [\textbf{Yes}/No/Not Applicable]
         \item A description of the computing infrastructure used. (e.g., type of GPUs, internal cluster, or cloud provider). [\textbf{Yes}/No/Not Applicable]
 \end{enumerate}

 \item If you are using existing assets (e.g., code, data, models) or curating/releasing new assets, check if you include:
 \begin{enumerate}
   \item Citations of the creator If your work uses existing assets. [\textbf{Yes}/No/Not Applicable]
   \item The license information of the assets, if applicable. [\textbf{Yes}/No/Not Applicable]
   \item New assets either in the supplemental material or as a URL, if applicable. [\textbf{Yes}/No/Not Applicable]
   \item Information about consent from data providers/curators. [Yes/No/\textbf{Not Applicable}]
   \item Discussion of sensible content if applicable, e.g., personally identifiable information or offensive content. [Yes/No/\textbf{Not Applicable}]
 \end{enumerate}

 \item If you used crowdsourcing or conducted research with human subjects, check if you include:
 \begin{enumerate}
   \item The full text of instructions given to participants and screenshots. [Yes/No/\textbf{Not Applicable}]
   \item Descriptions of potential participant risks, with links to Institutional Review Board (IRB) approvals if applicable. [Yes/No/\textbf{Not Applicable}]
   \item The estimated hourly wage paid to participants and the total amount spent on participant compensation. [Yes/No/\textbf{Not Applicable}]
 \end{enumerate}

 \end{enumerate}

\appendix

\onecolumn
\aistatstitle{Supplementary Materials}

\section{Details of experiments}\label{app:exp}

In this section, we provide some additional experiments and details on the 
experiments found in the main text of the paper. 

The license information for used assets is as follows:
\begin{itemize}
\item \textbf{Datasets}: We used the binary classification datasets from \url{https://jundongl.github.io/scikit-feature/datasets.html},
licensed under the GNU General Public License Version 2 (GPL-2.0).

\item \textbf{multiBUGS v2.0:} GNU Lesser General Public License Version 3 (LGPL-3.0).

\item \textbf{JAGS 4.3.2:} GNU General Public License Version 2.0 (GPL-2.0), MIT License.

\item \textbf{Stan NUTS 2.35.0:} The 3-Clause BSD License (BSD 3-clause).

\item \textbf{Julia package Pigeons:} GNU Affero General Public License Version 3 (AGPL-3.0).
\end{itemize}

\subsection{Details on \cref{sec:real_data_experiments}}\label{sec:real_data_details}
In this section, we give more details on the datasets as well as model specifics and 
data preprocessing procedures used in \cref{sec:experiments}.

\paragraph{Data details} We used the binary classification datasets from 
\url{https://jundongl.github.io/scikit-feature/datasets.html}, where a collection 
of classification datasets of varying sizes and sparsity is listed, along with their sources.
The datasets we used vary in size from $n=60$ to $n=7000$ observations, 
as well as from $d=500$ to $d=22283$ predictors. Across the datasets, 
$d/n$ ranges from $0.1927$ to $262.1647$. These details are summarized in \cref{tab:datasets}. 
Note that the sparsity is defined as the proportion of zero entries in the design matrix.

\paragraph{Priors} The priors we use in our models are as follows: 
\subparagraph{Gaussian prior} A Gaussian with mean 0 and standard deviation 10 is used for each parameter, i.e.
\[
  \theta_j\sim N(0,10^2), \quad \forall\ j\in\{1,\ldots,d\}.
\] 
\subparagraph{Horseshoe prior} A $t$-distribution prior is used for the intercept 
and zero mean Gaussian priors with standard deviation coming from a half-Cauchy 
are placed on the rest of the coefficients. That is,
\[
\theta_1 &\sim t(3,0,1) \\
\theta_j &\sim N(0,\lambda_j^2\tau^2), \quad \forall\ j\in\{2,\ldots,d\} \\
\lambda_j &\sim C^+(0,1), \quad \forall\ j\in\{2,\ldots,d\} \\
\tau &\sim C^+(0,1),
\] 
where $t(\nu,\mu,\sigma)$ denotes a t-distribution with $\nu$ degrees of freedom, 
location $\mu$, scale $\sigma$ and $C^+(0,1)$ denotes a half-Cauchy distribution.
See \cref{fig:horseshoe} in \cref{sec:HSP_example} for a visual representation of
the complex geometry induced by this prior.

\paragraph{Likelihood} We use a logistic regression likelihood, see 
\cref{sec:fast_gibbs}.

\begin{table}[t]
  \centering
  \caption{Datasets considered in our experiments.}
  \label{tab:datasets}
  \begin{tabular}{llll}
  \hline
  \textbf{Dataset} & \textbf{Sample size} & \textbf{Features} & \textbf{Sparsity}\\
  \hline
  \vspace{0.0cm}\\
  ALLAML & 72 & 7129 & 3.3$\times 10^{-5}$\\ 
  BASEHOCK & 1993 & 4862 & 0.9861\\ 
  GLI\textunderscore 85 & 85 & 22283 & 0\\ 
  PCMAC & 1943 & 3289 & 0.9854 \\ 
  Prostate\textunderscore GE & 102 & 5966 & 0 \\ 
  RELATHE & 1427 & 4322 & 0.9805 \\ 
  SMK\textunderscore CAN\textunderscore 187 & 187 & 19993 & 0\\ 
  arcene & 200 & 10000 & 0.4562\\
  colon & 62 & 2000 & 0.4158\\
  gisette & 7000 & 5000 & 0.87\\
  leukemia & 72 & 7070 & 0.4368\\
  madelon & 2600 & 500 & 7.6$\times 10^{-7}$ \vspace{0.25cm}\\
  \hline
  \end{tabular}
\end{table}

\subsection{CGGibbs algorithm}\label{sec:CGGibbs_algo}
The CGGibbs algorithm when applied to GLMs is stated in \cref{alg:CG_Gibbs}. 

\begin{algorithm}
	\begin{algorithmic}[1]
    \Require 
      Initial regression parameters $\beta^{(0)}$, 
      prior $\pi_0 = \otimes_{j=1}^d \pi_{0,j}$,
      observations $\cbra{(x_i,y_i)}_{i=1}^n$,
      inverse link function $\mu$, 
      response distribution $f_{Y|X}$ (mean parametrization), 
      \# MCMC iterations $T$, 
      conditional Markov kernel proposals $\cbra{Q_{j|-j}}_{j=1}^d$, 
      conditional targets $\cbra{\pi_{j | -j}}_{j=1}^d$

    \For{$i$ {\bf in} 1, 2, \dots, $n$} \Comment{One-time $O(dn)$ cache of linear predictors}
      \State $\texttt{cache}[i] \gets \sum_{j=1}^d \beta_j x_{ij}$ 
    \EndFor
		
    \State $\beta \gets \beta^{(0)}$
    \For{$t$ {\bf in} 1, 2, \dots, $T$}
      \For{$j$ {\bf in} 1, 2, \dots, $d$}
        \State $\beta_j' \sim Q_{j|-j}(\cdot | \beta_{-j})$ \Comment{Metropolis-within-Gibbs proposal}
        \State $\ell \gets 0$ 
        \For{$i$ {\bf in} 1, 2, \dots, $n$} \Comment{Increment log-likelihood}
          \State $\texttt{linear\_predictor} \gets \texttt{cache}[i] - \beta_j x_{ij} + \beta_j' x_{ij}$
          \State $\ell \gets \ell + \log(f_{Y|X}(y_i | \mu(\texttt{linear\_predictor})))$
	    \EndFor
      \State $U \gets \text{Unif}(0,1)$ 
      \State $\alpha \gets \texttt{MH\_probability}(\ell, \pi_{j | -j}, Q(\cdot | \beta_{-j}), \pi_{0,j})$
        \Comment{Acceptance probability}
      \If{$U \leq \alpha$} \Comment{Accept proposal}
        \For{$i$ {\bf in} 1, 2, \dots, $n$} \Comment{$O(n)$ update of cache}
          \State $\texttt{cache}[i] \gets \texttt{cache}[i] - \beta_j x_{ij} + \beta_j' x_{ij}$ 
        \EndFor 
        \State $\beta_j \gets \beta_j'$ \Comment{Update parameter}
      \EndIf
	  \EndFor
      \State $\beta^{(t)} \gets \beta$ 
	\EndFor
    \State \Return $\{\beta^{(t)}\}_{t=0}^T$
	\end{algorithmic}
  \caption{Compute graph Gibbs for GLMs ($T$ passes over all parameters)}
  \label{alg:CG_Gibbs}
\end{algorithm}

\subsection{Implementation details}
All MCMC chains are run until a minimum ESS 
of 100 is reached. Here both CGGibbs and Stan NUTS use half of the total number 
of iterations as warmup and warmup samples are not used in the ESS calculations. 
The reasoning for why the ESS provides an adequate estimate of divergence scalings 
is given in \cref{sec:convrate_vs_ESS}. 
We note that in our simulations, when subsampling is performed for each replicate,
we change the shuffling of the data as well as the seed for the MCMC chains.

\subparagraph{CGGibbs} We use the slice sampler \cite{neal_slice_2003} from the Julia 
package \verb|Pigeons| \citep{surjanovic2023pigeons} 
and set the number of passes through all variables 
per exploration step as 1 while the other hyperparameters are set to their default values.

\subparagraph{Stan NUTS} We use Stan NUTS (2.35.0) \cite{carpenter2017stan} and set all 
hyperparameters other than warmup time to their defaults. 

\paragraph{Data preprocessing} We follow the standard practice of standardizing the design 
matrix of each dataset before supplying them to the samplers. Surprisingly, we 
observe that Stan NUTS benefits from sparse design matrices (sparsity greater than 0.85) even 
without sparse encoding in the model. Therefore, we opt rescale the non-zero entries of sparse datasets
by dividing by their maximum absolute values for each column 
for both Stan NUTS and CGGibbs, making the comparison as fair as possible.

\subsection{More dimensional scaling results}\label{sec:more_dim_scale}
\begin{figure}[t]
  \centering
  \begin{subfigure}{0.3\textwidth}
    \centering
    \includegraphics[width=\textwidth]{img/dim_scale_ex/dim_scale_medESS_colon.png}
  \end{subfigure}
  \begin{subfigure}{0.3\textwidth}
    \centering
    \includegraphics[width=\textwidth]{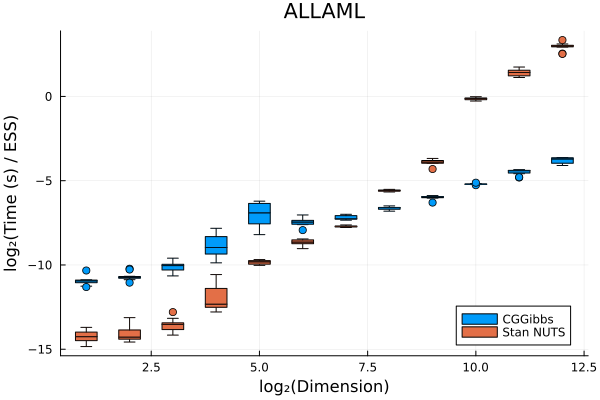}
  \end{subfigure}
  \begin{subfigure}{0.3\textwidth}
    \centering
    \includegraphics[width=\textwidth]{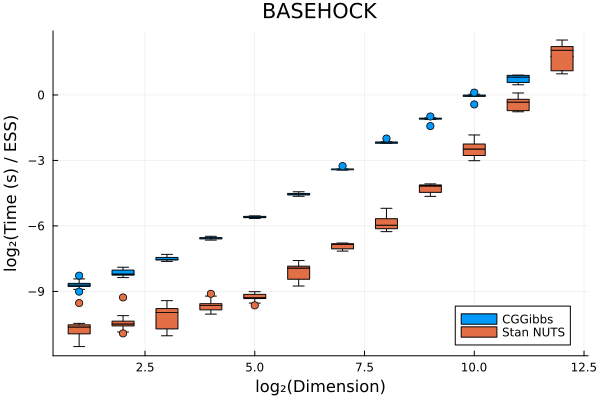}
  \end{subfigure}\\
  \begin{subfigure}{0.3\textwidth}
    \centering
    \includegraphics[width=\textwidth]{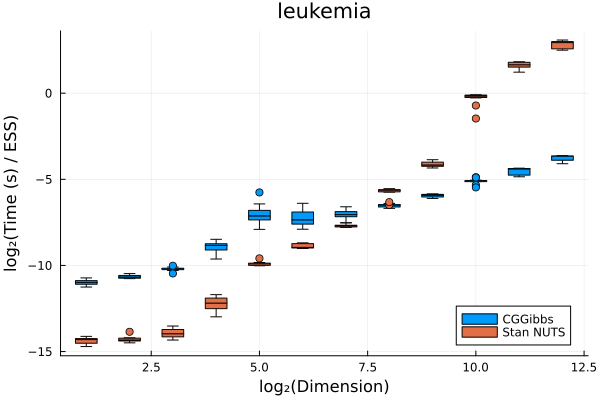}
  \end{subfigure}
  \begin{subfigure}{0.3\textwidth}
    \centering
    \includegraphics[width=\textwidth]{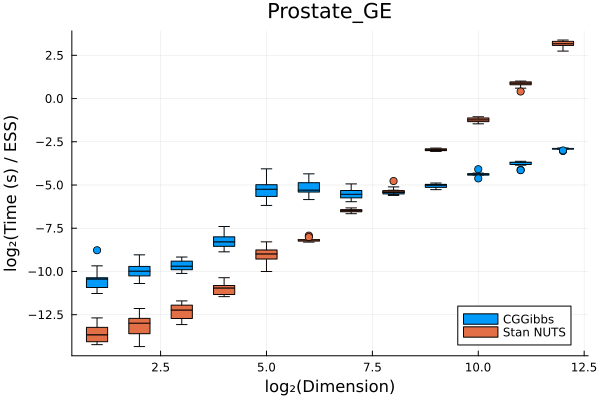}
  \end{subfigure}
  \begin{subfigure}{0.3\textwidth}
    \centering
    \includegraphics[width=\textwidth]{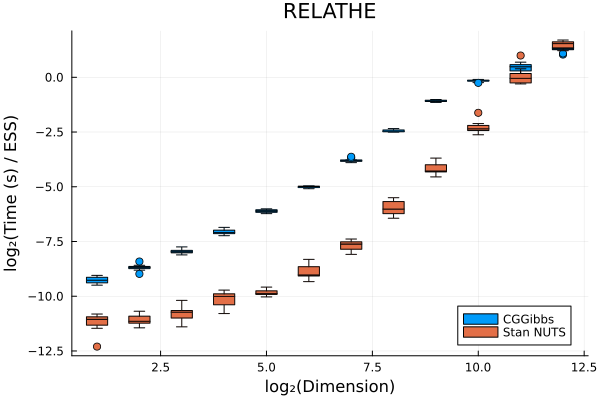}
  \end{subfigure}\\
  \begin{subfigure}{0.3\textwidth}
    \centering
    \includegraphics[width=\textwidth]{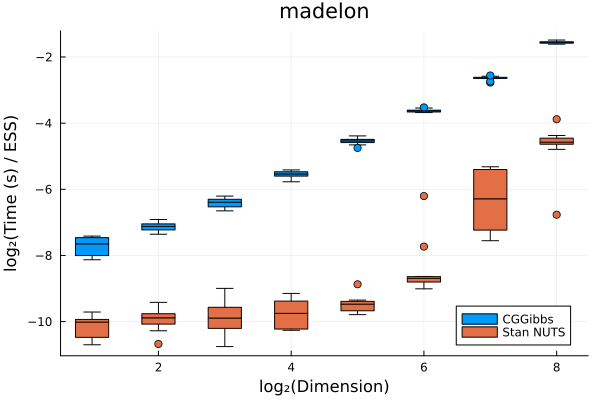}
  \end{subfigure}
  \begin{subfigure}{0.3\textwidth}
    \centering
    \includegraphics[width=\textwidth]{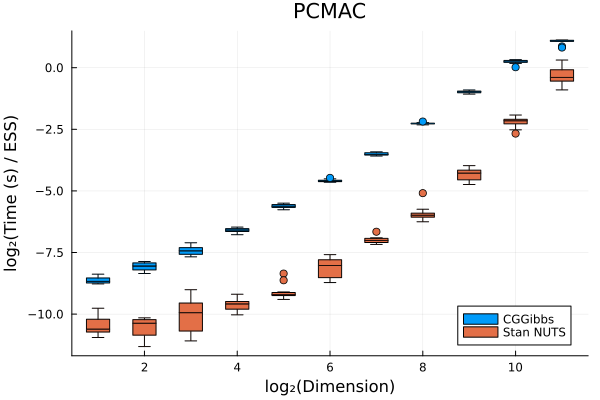}
  \end{subfigure}
  \caption{Time per median ESS for CGGibbs
  and Stan NUTS as a function of dimension for several other datasets. 
  The box plots summarize the results over 10 replicates.}
  \label{fig:dim_scale_other_data_med}
\end{figure}

\begin{figure}[t]
  \centering
  \begin{subfigure}{0.3\textwidth}
    \centering
    \includegraphics[width=\textwidth]{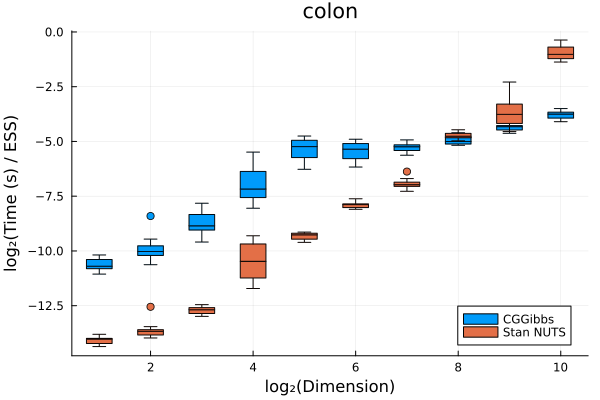}
  \end{subfigure}
  \begin{subfigure}{0.3\textwidth}
    \centering
    \includegraphics[width=\textwidth]{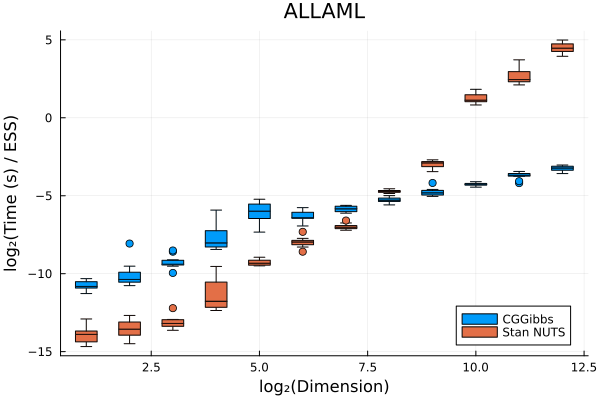}
  \end{subfigure}
  \begin{subfigure}{0.3\textwidth}
    \centering
    \includegraphics[width=\textwidth]{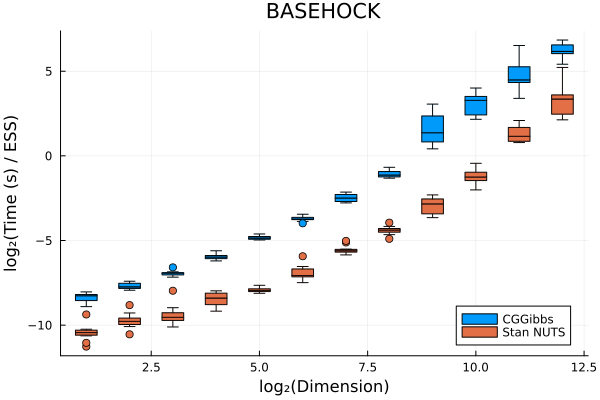}
  \end{subfigure}\\
  \begin{subfigure}{0.3\textwidth}
    \centering
    \includegraphics[width=\textwidth]{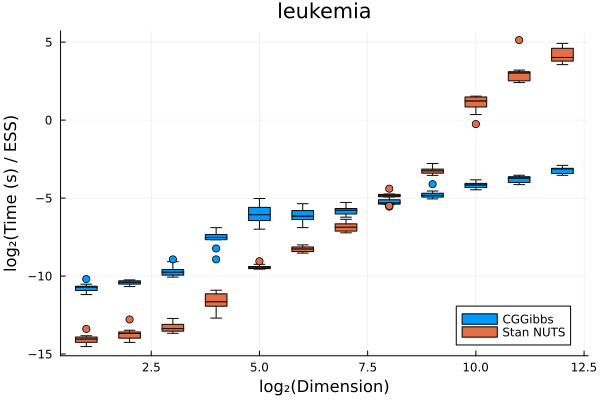}
  \end{subfigure}
  \begin{subfigure}{0.3\textwidth}
    \centering
    \includegraphics[width=\textwidth]{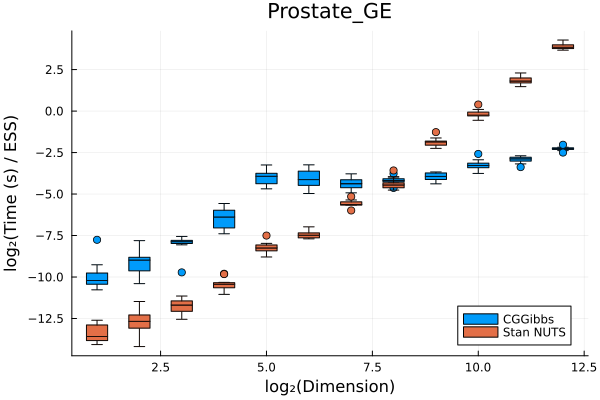}
  \end{subfigure}
  \begin{subfigure}{0.3\textwidth}
    \centering
    \includegraphics[width=\textwidth]{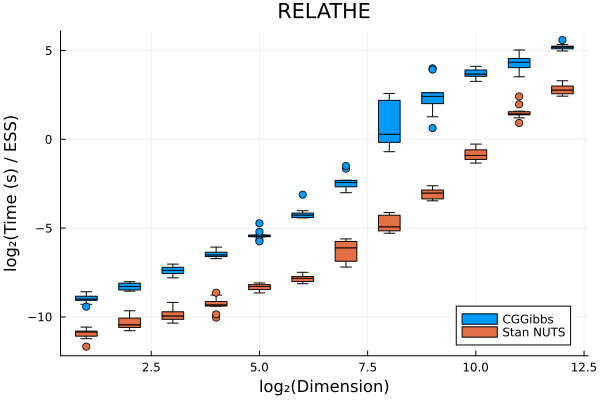}
  \end{subfigure}\\
  \begin{subfigure}{0.3\textwidth}
    \centering
    \includegraphics[width=\textwidth]{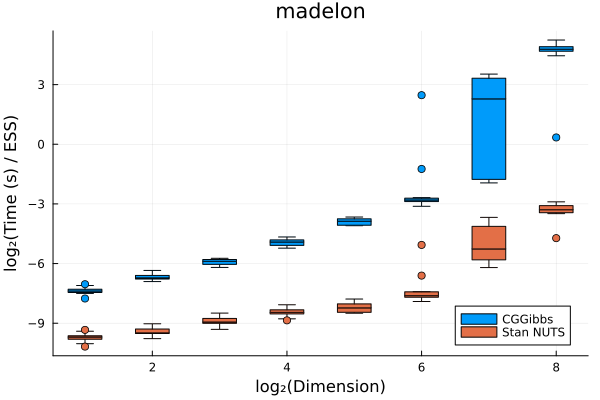}
  \end{subfigure}
  \begin{subfigure}{0.3\textwidth}
    \centering
    \includegraphics[width=\textwidth]{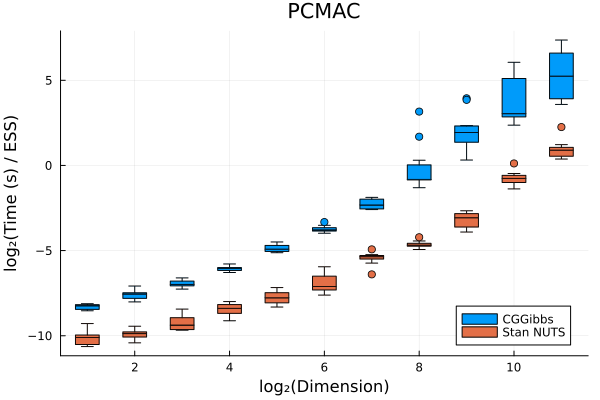}
  \end{subfigure}
  \caption{Time per min ESS for CGGibbs
  and Stan NUTS as a function of dimension for several other datasets.
  The box plots summarize the results over 10 replicates.}
  \label{fig:dim_scale_other_data_min}
\end{figure}

In this section, we repeat the increasing column experiment previously done for the colon cancer 
dataset on other datasets. These are datasets from \cref{sec:real_data_experiments} where 
both CGGibbs and Stan NUTS achieve a minimum ESS of 100 within 3 days using a Gaussian prior. 
For each dataset, there are 10 replicates runs with a different shuffling 
of the features and a different random seed for the MCMC chain. The results are shown in 
\cref{fig:dim_scale_other_data_med,fig:dim_scale_other_data_min}.

\begin{figure}[t]
  \centering
  \begin{subfigure}{0.3\textwidth}
    \centering
    \includegraphics[width=\textwidth]{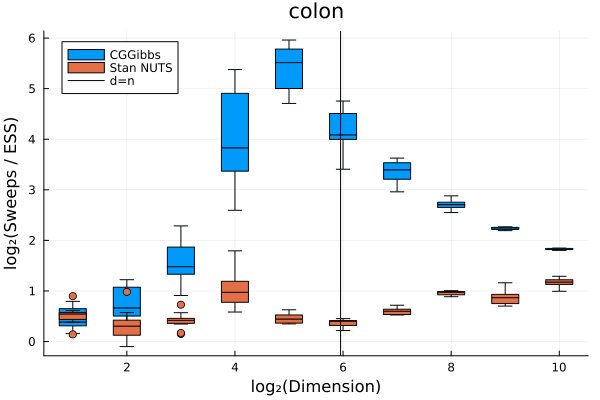}
  \end{subfigure}
  \begin{subfigure}{0.3\textwidth}
    \centering
    \includegraphics[width=\textwidth]{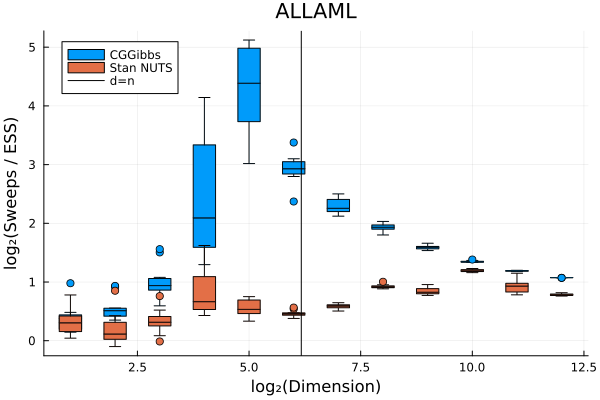}
  \end{subfigure}
  \begin{subfigure}{0.3\textwidth}
    \centering
    \includegraphics[width=\textwidth]{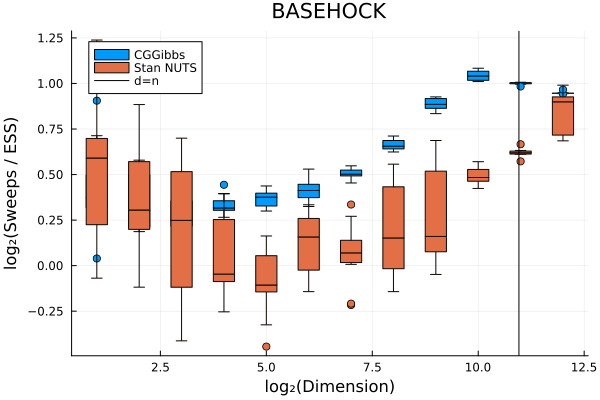}
  \end{subfigure}\\
  \begin{subfigure}{0.3\textwidth}
    \centering
    \includegraphics[width=\textwidth]{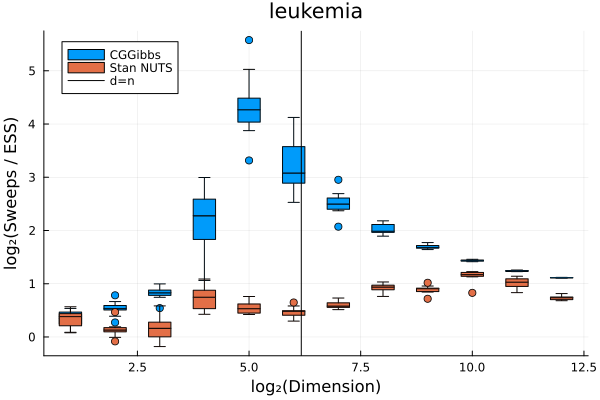}
  \end{subfigure}
  \begin{subfigure}{0.3\textwidth}
    \centering
    \includegraphics[width=\textwidth]{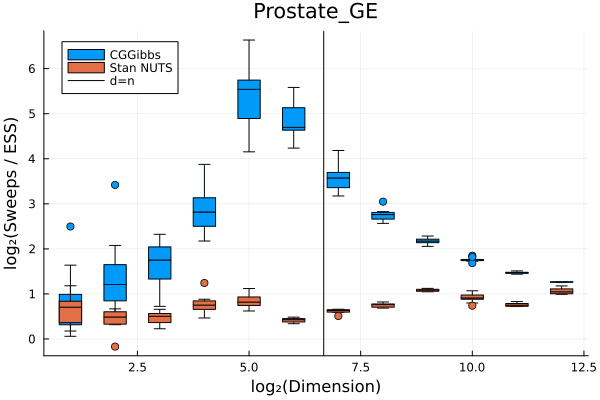}
  \end{subfigure}
  \begin{subfigure}{0.3\textwidth}
    \centering
    \includegraphics[width=\textwidth]{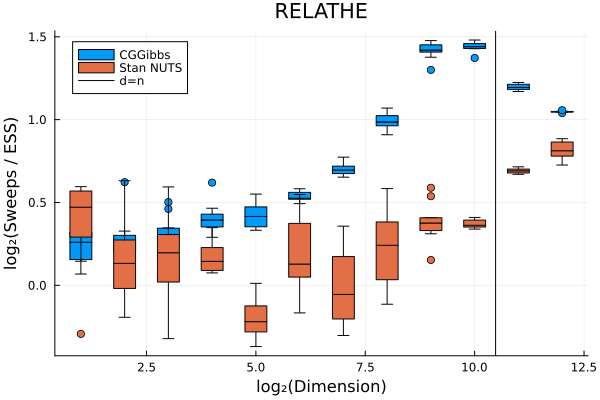}
  \end{subfigure}\\
  \begin{subfigure}{0.3\textwidth}
    \centering
    \includegraphics[width=\textwidth]{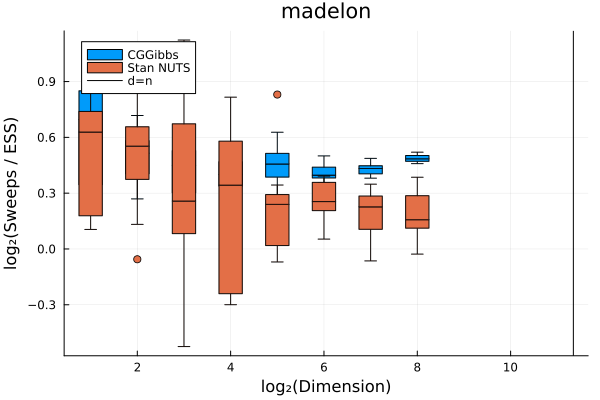}
  \end{subfigure}
  \begin{subfigure}{0.3\textwidth}
    \centering
    \includegraphics[width=\textwidth]{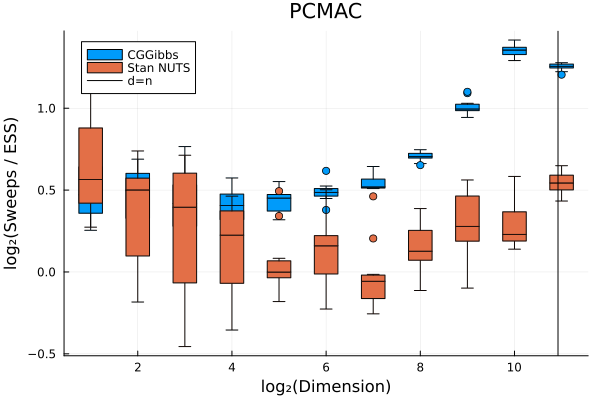}
  \end{subfigure}
  \caption{Sweeps per median ESS for CGGibbs
  and Stan NUTS as a function of dimension. The black line indicates the point at which $d=n$.
  The box plots summarize the results across 10 replicates.}
  \label{fig:sweeps_to_medESS}
\end{figure}

\begin{figure}[t]
  \centering
  \begin{subfigure}{0.3\textwidth}
    \centering
    \includegraphics[width=\textwidth]{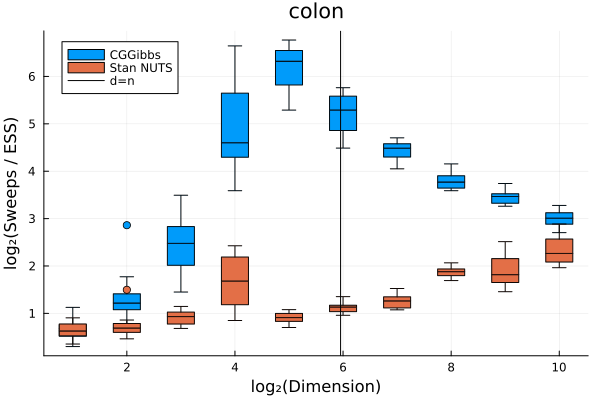}
  \end{subfigure}
  \begin{subfigure}{0.3\textwidth}
    \centering
    \includegraphics[width=\textwidth]{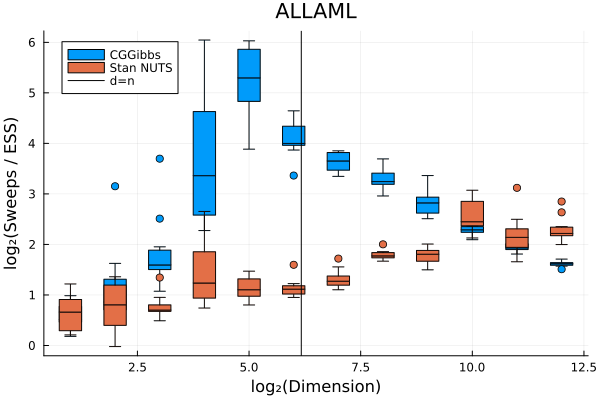}
  \end{subfigure}
  \begin{subfigure}{0.3\textwidth}
    \centering
    \includegraphics[width=\textwidth]{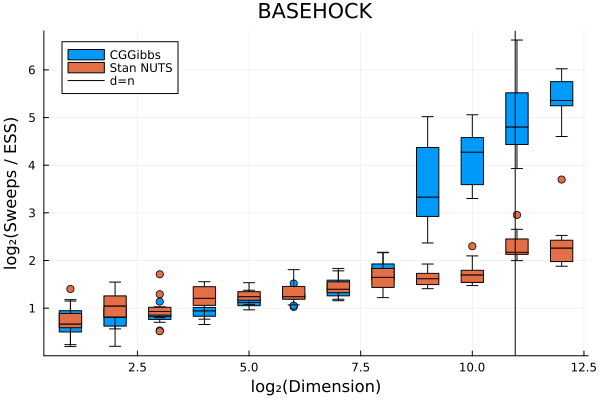}
  \end{subfigure}\\
  \begin{subfigure}{0.3\textwidth}
    \centering
    \includegraphics[width=\textwidth]{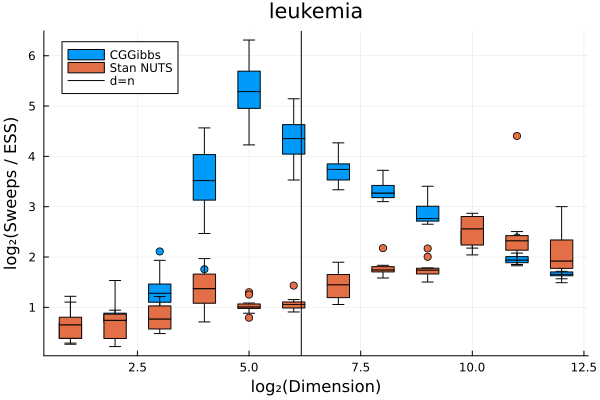}
  \end{subfigure}
  \begin{subfigure}{0.3\textwidth}
    \centering
    \includegraphics[width=\textwidth]{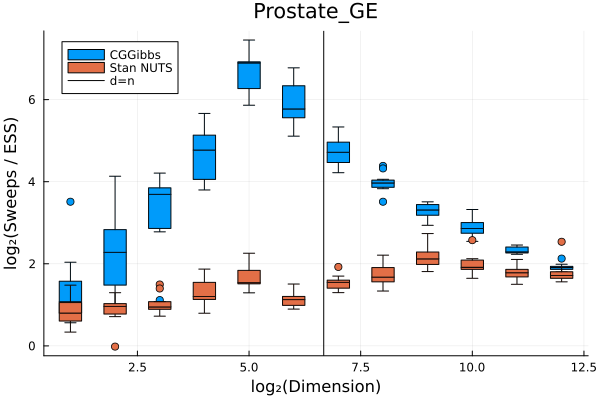}
  \end{subfigure}
  \begin{subfigure}{0.3\textwidth}
    \centering
    \includegraphics[width=\textwidth]{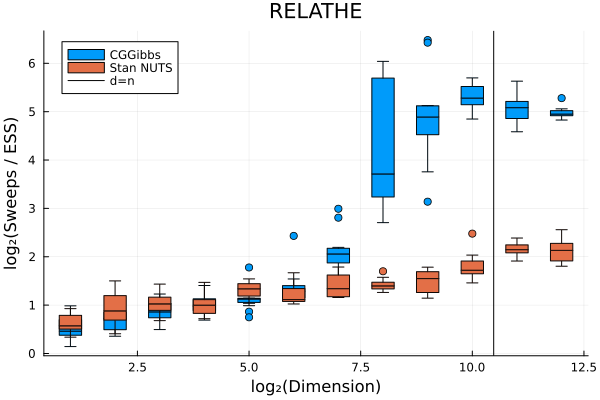}
  \end{subfigure}\\
  \begin{subfigure}{0.3\textwidth}
    \centering
    \includegraphics[width=\textwidth]{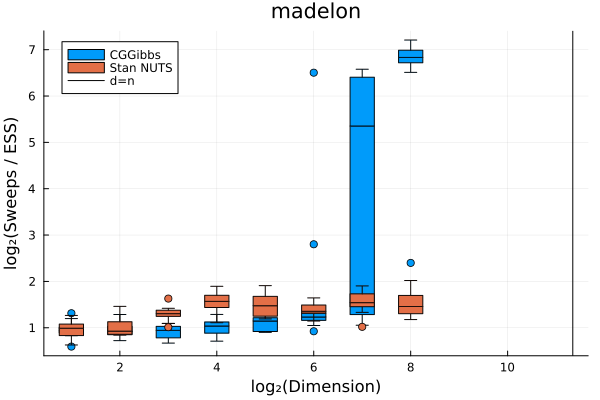}
  \end{subfigure}
  \begin{subfigure}{0.3\textwidth}
    \centering
    \includegraphics[width=\textwidth]{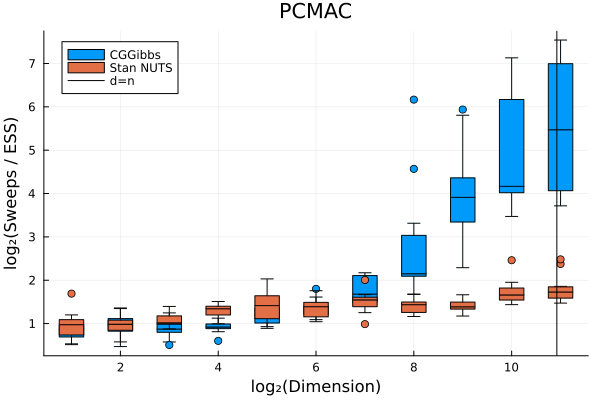}
  \end{subfigure}
  \caption{Number of sweeps per min ESS for CGGibbs
  and Stan NUTS as a function of dimension. The black line indicates the point at which $d=n$.
  The box plots summarize the results across 10 replicates.}
  \label{fig:sweeps_to_minESS}
\end{figure}

In addition, we also show the number of sweeps (or iterations for NUTS) per 
median ESS in \cref{fig:sweeps_to_medESS,fig:sweeps_to_minESS}, where the shift 
from $d<n$ to $d>n$ is more evident. 
Observe that this shift does not always occur at the point where $d=n$ but usually in its vicinity.

\subsection{Condition number scaling}\label{sec:cond_num_scale}
\begin{figure}[t]
  \centering
  \begin{subfigure}{0.3\textwidth}
    \centering
    \includegraphics[width=\textwidth]{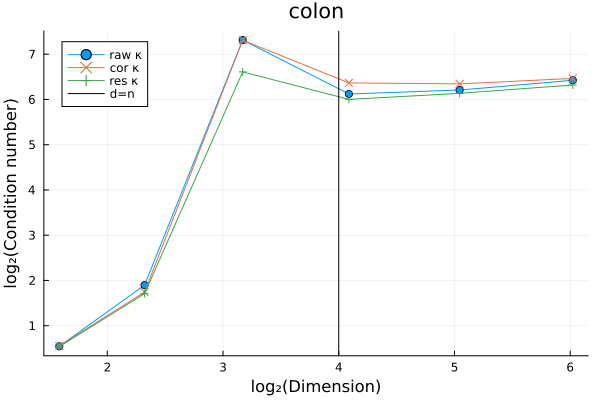}
  \end{subfigure}
  \begin{subfigure}{0.3\textwidth}
    \centering
    \includegraphics[width=\textwidth]{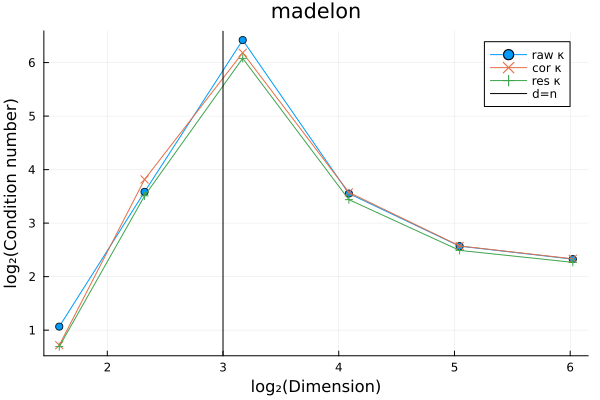}
  \end{subfigure}
  \begin{subfigure}{0.3\textwidth}
    \centering
    \includegraphics[width=\textwidth]{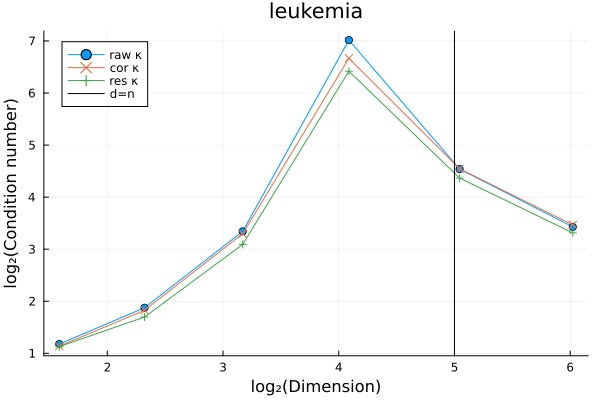}
  \end{subfigure}\\
  \begin{subfigure}{0.3\textwidth}
    \centering
    \includegraphics[width=\textwidth]{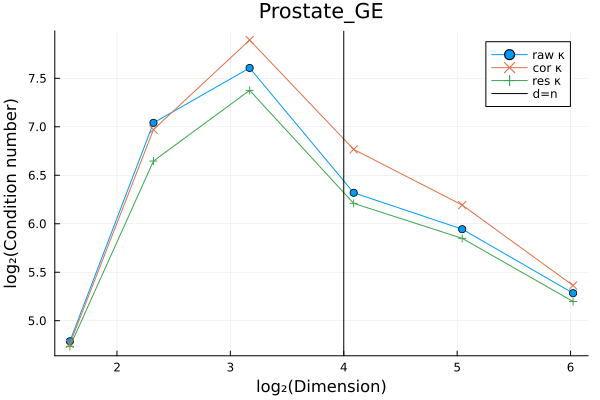}
  \end{subfigure}
  \begin{subfigure}{0.3\textwidth}
    \centering
    \includegraphics[width=\textwidth]{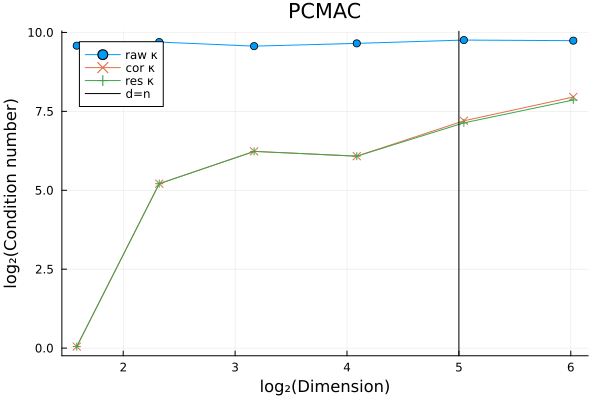}    
  \end{subfigure}
  \caption{Dimensional scaling of various notions of condition number 
  when the Gaussian prior is used for different datasets and sample sizes. 
  The black line indicates the point at which $d=n$.}
  \label{fig:cond_num_scale}
\end{figure}

In this section, we show several examples of the shift in sampling efficiency 
from $d<n$ to $d>n$ also occuring for various 
notions of condition number when the Gaussian prior is used. The results are shown 
in \cref{fig:cond_num_scale}. Based on this figure, we conclude that the shift 
in sampling efficiency may not occur for sparse data such as the 
PCMAC newsgroup dataset. 

\subsection{Impact of irrelevant parameters on the Gibbs mixing rate}\label{sec:irrelevant_params}
\begin{figure}[t]
  \centering
  \begin{subfigure}{0.48\textwidth}
    \centering
    \includegraphics[width=\textwidth]{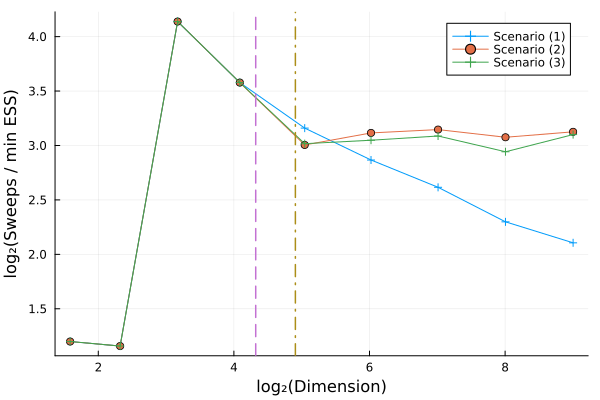}
  \end{subfigure}
  \begin{subfigure}{0.48\textwidth}
    \centering
    \includegraphics[width=\textwidth]{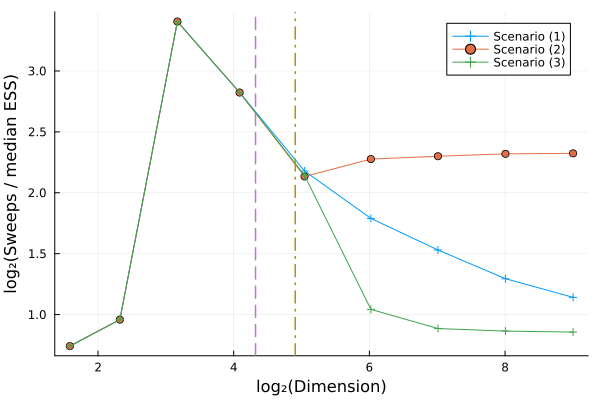}    
  \end{subfigure}
  \caption{Effect of extra features on the number of Gibbs 
  sweeps per ESS when the Gaussian prior is used in three different scenarios: 
  (1) no multicollinearity in the design matrix, (2) multicollinearity present in the design matrix, 
  and (3) irrelevant features are identically zero. 
  The dashed line is where $d=n$ and the dash-dotted line is where 
  the ``irrelevant features'' start to be added. \textbf{Left:} results in terms of sweeps per min ESS, 
  \textbf{reft:} results in terms of sweeps per median ESS}
  \label{fig:extra_features}
\end{figure}

To further examine the effect of varying the number of features on the Gibbs sampler, as seen
in \cref{sec:empirical-scaling}, we generate three synthetic 
datasets where only a prefix of the features influence the outcome. 
Specifically, each dataset has 
\begin{align}
  \text{Design matrix:}&\quad x = (x_1, x_2, \ldots, x_n)^\top\in \reals^{n\times d},\\
  \text{True parameter:}&\quad \theta = (a,\theta_s,\textbf{0}) \in \reals^d,\\
  \text{Logistic outcome:}&\quad y_i \mid x_i \sim \Bern(\text{logistic}(a+\theta^\top x_i)),\quad i=1,\ldots,n,
\end{align}
where $d=2^{15}, n=20, x_i \in \reals^d$ for all $i=1,\ldots,n$, $a\in \reals$ is the intercept and 
$\theta_s\in \reals^{30}$ is the coefficient vector for the significant features. 
The design matrices for each scenario are as follows:
\begin{enumerate}
  \item[(1)] Set $x_i\sim \Norm(0,I_d)$ for all $i=1,\ldots,n$. In this setting, 
  the features are uncorrelated.
  \item[(2)] Set $x_i$ the same as scenario (1) for all $i\le 30$ and set $x_i=x_1$ for all $i>30$. 
  Here, extra features 
  are perfectly correlated with the first (significant) feature.
  \item[(3)] Set $x_i$ the same as scenario (1) for all $i\le 30$ and set $x_i=0$ for all $i>30$. 
\end{enumerate}
We use CGGibbs to sample from these synthetic datasets using a Gaussian prior.
The results of these experiments are shown in \cref{fig:extra_features}. 
The results confirm the surprising findings of \cref{sec:empirical-scaling} that adding more features can help the Gibbs sampler even if 
the added features are not related to the data generation process---depending on 
how ``well behaved" the extra features are. In the case of an overparameterized GLM,
the irrelevant dimensions might be related to the auxiliary variables used in data augmentation samplers 
\citep{tanner1987calculation}. We suspect that the extra features let the 
parameters move more freely in a larger state space, thereby speeding up the 
overall mixing of the chain. 
Based on these results, it seems that the Gibbs sampler can potentially benefit from overparameterization.

In addition, zero elements can decrease the number of sweeps per median ESS while not helping 
decrease the number of sweeps per min ESS. This is because updates to components associated with 
zero elements are not affected by the likelihood, thereby making the chains of these components 
sample from the priors. As prior distributions are typically easy to sample from, the ESS of 
these components can increase, which increases the median ESS across 
marginals (but not the min ESS, which measures the most difficult marginal). 
We also observe this disparity between min and median ESS in sparse datasets in 
\cref{sec:real_data_experiments_min}

\subsection{Condition number adaptation with NUTS}\label{sec:cond_num_speed}
\begin{figure}[t]
  \centering
  \includegraphics[width=0.48\textwidth]{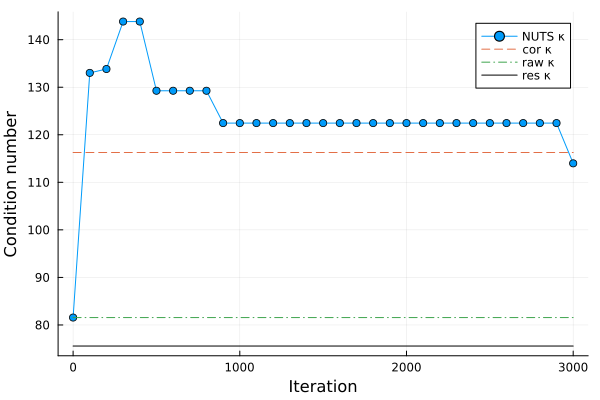}
  \caption{Condition number (lower is better) adaption used by NUTS for a 
  logistic regression problem with a Gaussian prior. The condition number is plotted 
  as a function of the number of MCMC iterations. 
  The green dashed line is the initial condition number (no preconditioning), the red dashed line 
  is the condition number after preconditioning with marginal standard deviations and the solid 
  black line is the best linear preconditioner. 
  The NUTS condition number (blue solid line) eventually converges to the red dashed line.}
  \label{fig:diag_cond_speed}
\end{figure}

In this section, we show an example of a problem where $\kappac<\kappa$. 
For this problem, we apply a logistic regression model with a Gaussian prior on a subset of the
prostate cancer dataset where $d=n=16$ and record the progression of 
the condition number adaptation in Stan NUTS. The result is shown in \cref{fig:diag_cond_speed}.

\subsection{Horseshoe prior example}\label{sec:HSP_example}

\begin{figure}[t]
\centering
\includegraphics[width=0.6\textwidth]{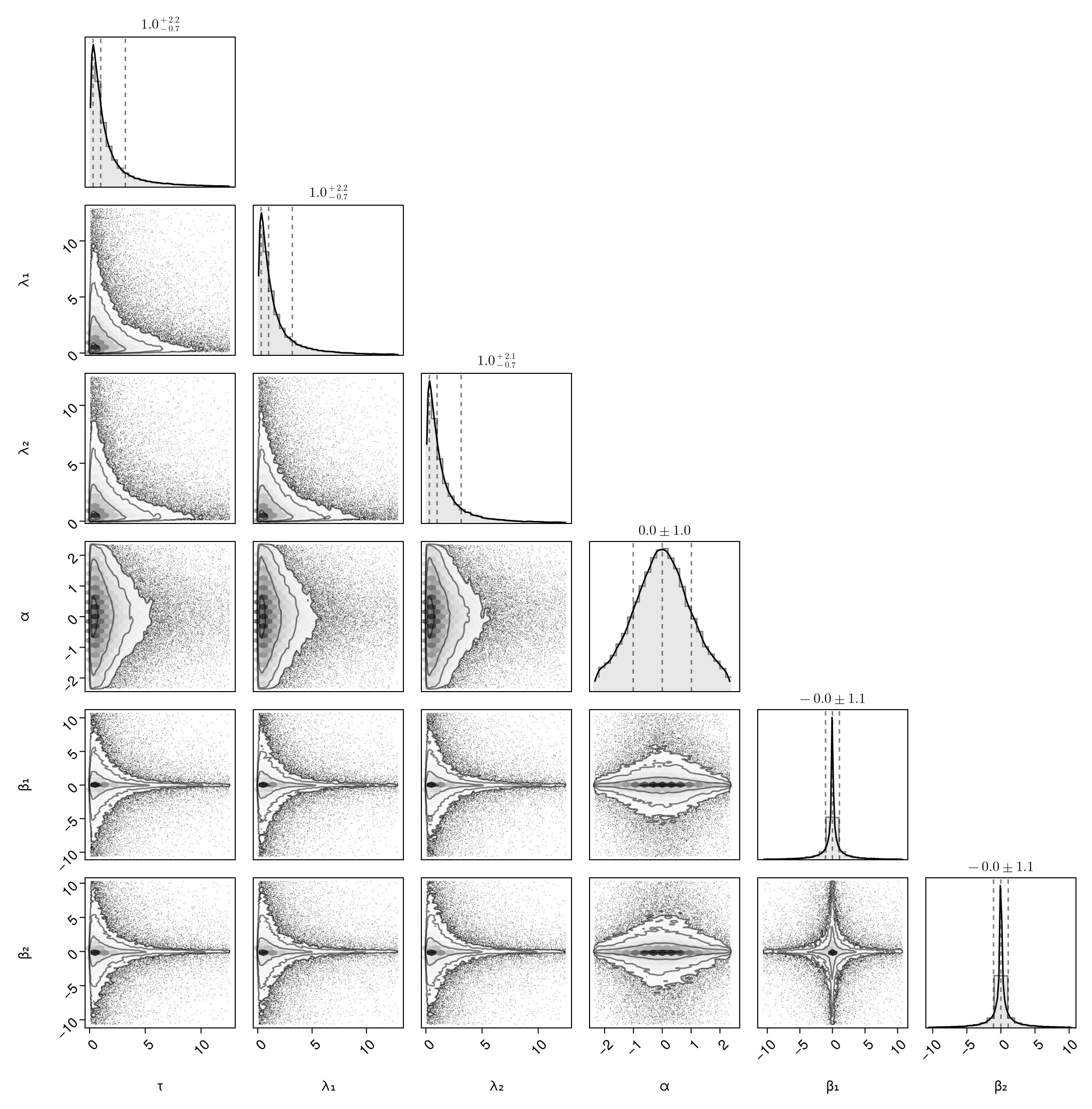}
\caption{Univariate and bivariate marginals of the horseshoe prior for 
$\beta\in\reals^2$, approximated by drawing 100,000 \iid samples from the generative model.}
\label{fig:horseshoe}
\end{figure}

\cref{fig:horseshoe} shows a visual representation of the horseshoe prior for the 
case $\beta\in\reals^2$. Notice that, while the univariate marginals would suggest a 
log-concave distribution, the bivariate densities indicate a much more complex
geometry. 

\subsection{Panel of datasets: results for min ESS}\label{sec:real_data_experiments_min}
\begin{figure}[t]
  \centering
  \begin{subfigure}{0.4\textwidth}
    \centering
    \includegraphics[width=\textwidth]{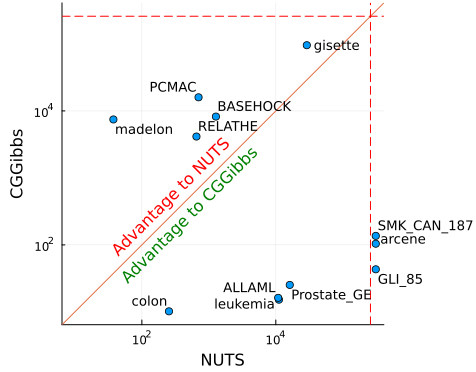} 
    \caption{Gaussian prior}  
  \end{subfigure}
  \begin{subfigure}{0.4\textwidth}
    \centering
    \includegraphics[width=\textwidth]{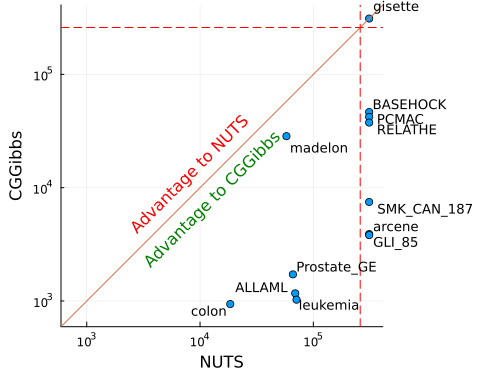}    
    \caption{Horseshoe prior}
  \end{subfigure}
  \caption{Median (across 30 replicates) wall-clock time (in seconds) per 100 minimum 
  ESS for CGGibbs and Stan NUTS 
  on \ndatasetstotal\ logistic regression problems with real datasets. We use both a Gaussian prior (left) 
  and horseshoe prior (right). 
  Each point is a dataset: its x-axis coordinate denotes the median time for NUTS, and its y-axis 
  coordinate denotes the median time for CGGibbs. The region beyond 
  the dashed lines indicates that the corresponding sampler did not reach a minimum of 100 
  ESS within three days.}
  \label{fig:real_data_ess_min}
\end{figure}

The time per minimum ESS results for the panel of datasets considered in 
\cref{sec:real_data_experiments} are shown in 
\cref{fig:real_data_ess_min}. 
For non-sparse datasets, the performance for both samplers in terms of the minimum ESS
is similar to the median ESS counterpart. For sparse data, however, Stan 
NUTS performs better in terms of time per minimum ESS. This behaviour is 
in line with the results from \cref{sec:irrelevant_params} as a sparse data 
matrix can be interpreted as being similar to the third scenario in that section. 

\section{Proofs} 
In this section we prove \cref{thm:Gibbs_convergence} and \cref{prop:conditioning_invariance_Gibbs}.

\subsection{Proof of \cref{thm:Gibbs_convergence}}
\bprfof{\cref{thm:Gibbs_convergence}}
To establish the convergence rate of DUGS, 
we first define some relevant matrices. Let 
\[
  \Sigma = 
  \begin{pmatrix}
    \Sigma_{11} & \Sigma_{12} & \ldots & \Sigma_{1d} \\
    \Sigma_{21} & \Sigma_{22} & \ldots & \Sigma_{2d} \\
    \vdots      & \vdots     & \ddots & \vdots \\
    \Sigma_{d1} & \Sigma_{d2} & \ldots & \Sigma_{dd}
  \end{pmatrix}, 
  \qquad
  \Sigma^{-1} := Q = 
  \begin{pmatrix}
    Q_{11} & Q_{12} & \ldots & Q_{1d} \\
    Q_{21} & Q_{22} & \ldots & Q_{2d} \\
    \vdots & \vdots & \ddots & \vdots \\
    Q_{d1} & Q_{d2} & \ldots & Q_{dd}
  \end{pmatrix}. 
\]
Define 
\[
  A(\Sigma) = I - \diag(Q_{11}^{-1}, Q_{22}^{-1}, \ldots, Q_{dd}^{-1}) Q, 
\]
and let $L(\Sigma)$ be a block lower triangular matrix such that the lower triangle blocks 
coincide with those of $A(\Sigma)$. Finally, set $U(\Sigma) = A(\Sigma)-L(\Sigma)$ and define 
\[
  B(\Sigma) = (I-L(\Sigma))^{-1} U(\Sigma).
\]
We express the convergence rate for each metric in terms of the spectral radius of the matrix $B(\Sigma)$, 
denoted $\rho(B(\Sigma))$, which is the modulus of the eigenvalue of $B(\Sigma)$ with the largest modulus.
From \cref{lem:rho_B_Sigma}, we have that 
\[
  \rho(B(\Sigma)) \leq \exp\left(-\frac{1}{\kappa(\Sigma)}\right).
\]

\paragraph{TV bound.} First, we prove that the rate of convergence of DUGS in TV distance is 
given by $\rho(B(\Sigma))$. From Theorem 1 of \cite{roberts1997updating}, we have 
$\pi_t := \DUGS^t(x, \cdot) = \Norm(\mu_t,\Sigma_t)$ for all $t\ge 0$, where
\[ \label{eq:gibbsgaussian}
    \mu_t &= \mu + B(\Sigma)^t(\mu_0-\mu), \qquad
    \Sigma_t = \Sigma + B(\Sigma)^t(\Sigma_0-\Sigma)(B(\Sigma)^\top)^t
\]
with $\Norm(\mu_0,\Sigma_0)$ being the initial distribution. 
(I.e., here $\mu_0 = x$ and $\Sigma_0 = 0$.)
By Pinsker's inequality, we can bound the total variation with the KL divergence,  
and the KL divergence between two Gaussians has a closed-form expression, so that
\[
    \TV(\pi, \pi_t) 
    &\leq \sqrt{\frac{1}{2}\KL(\pi || \pi_t)}\\
\label{eq:pinsker}
    &= \frac{1}{\sqrt{2}} \sqrt{
        \tr(\Sigma^{-1}\Sigma_t - I) + (\mu - \mu_t)^\top \Sigma^{-1} (\mu - \mu_t) - \log \det (\Sigma_t \Sigma^{-1}).
    }
\]
We analyze the convergence for each of the terms inside the square root. 
By \cref{eq:gibbsgaussian},
\[\label{eq:tvterm1}
    \tr(\Sigma^{-1}\Sigma_t - I) &= \tr\left( \Sigma^{-1} (\Sigma + B(\Sigma)^t(\Sigma_0 - \Sigma) (B(\Sigma)^t)^\top) - I\right) \\
                                 &= \tr(\Sigma^{-1}B(\Sigma)^t(\Sigma_0 - \Sigma) (B(\Sigma)^t)^\top) \\
                                 &= O(\rho(B(\Sigma))^{2t}),
\]
where the asymptotic rate is due to the fact that $B(\Sigma)^t$ converges to 0 
element-wise at rate $\rho(B(\Sigma))$ \cite[Lemma 4]{roberts1997updating}. 
Similarly, combining \cref{eq:gibbsgaussian} and the convergence of $B(\Sigma)^t$, we have that 
\[
  \|\mu - \mu_t\| = \|B^t (\mu - \mu_0)\| = O(\rho(B(\Sigma))^t), 
\]
yielding 
\[\label{eq:tvterm2}
    (\mu - \mu_t)^\top \Sigma^{-1} (\mu - \mu_t) = O(\rho(B(\Sigma))^{2t}).
\]
For the third term, we obtain that 
\[\label{eq:tvterm3}
    \abs{\log \det (\Sigma_t \Sigma^{-1})} &= \abs{\log \det (I + B(\Sigma)^t(\Sigma_0 - \Sigma) (B(\Sigma)^t)^\top\Sigma^{-1})}\\
                                           &\leq \tr(B(\Sigma)^t(\Sigma_0 - \Sigma) (B(\Sigma)^t)^\top \Sigma^{-1})\\
                                           &= O(\rho(B(\Sigma))^{2t}).
\]
Finally, combining \cref{eq:pinsker} and \cref{eq:tvterm1,eq:tvterm2,eq:tvterm3} yields 
\[ \label{eq:TV_convergence}
  \TV(\pi, \pi_t) = O(\rho(B(\Sigma))^t).
\]

\paragraph{Wasserstein bounds.} Next, we prove the Wasserstein bounds.
For two Gaussian distributions $\pi_1 = \Norm(\mu_1,\Sigma_1)$ and 
$\pi_2 = \Norm(\mu_2,\Sigma_2)$, we have the following formula for 
their Wasserstein 2-distance:
\begin{align}\label{wass2_normal}
    \Wass_2^2(\pi_1,\pi_2) = \|\mu_1-\mu_2\|_2^2 + \tr(\Sigma_1+\Sigma_2
    - 2(\Sigma_2^{1/2}\Sigma_1\Sigma_2^{1/2})^{1/2}),
\end{align}
where $\tr(M)$ is the trace of the matrix $M$ and 
$M^{1/2}$ is the principal square root of $M$.
Applying \eqref{wass2_normal} to $\pi_1 = \pi_t = \Norm(\mu_t,\Sigma_t)$ and 
$\pi_2 = \pi = \Norm(\mu,\Sigma)$ gives us
\begin{align}\label{wass2_init}
    \Wass_2^2(\pi,\pi_t) &= \|\mu-\mu_t\|_2^2 + \tr(\Sigma+\Sigma_t
    - 2(\Sigma^{1/2}\Sigma_t\Sigma^{1/2})^{1/2})\\
    &= \|B(\Sigma)^t(\mu_0-\mu)\|_2^2 + \tr(2\Sigma+B(\Sigma)^t(\Sigma_0-\Sigma)(B(\Sigma)^\top)^t)\\
    &\qquad\qquad - 2\tr(\Sigma^{1/2}(\Sigma + B(\Sigma)^t(\Sigma_0-\Sigma)(B(\Sigma)^\top)^t)\Sigma^{1/2})^{1/2})\\
    &=\|B(\Sigma)^t(\mu_0-\mu)\|_2^2 + 2\tr(\Sigma)+ \tr(B(\Sigma)^t(\Sigma_0-\Sigma)(B(\Sigma)^\top)^t)\\
    &\qquad\qquad - 2\tr((\Sigma^{2} + (\Sigma^{1/2}B(\Sigma)^t(\Sigma_0-\Sigma)(B(\Sigma)^\top)^t\Sigma^{1/2})^{1/2})\\
    &=\|B(\Sigma)^t(\mu_0-\mu)\|_2^2 + 2\tr(\Sigma)+ \tr(M_t) - 2\tr((\Sigma^{2} + (\Sigma^{1/2}M_t\Sigma^{1/2})^{1/2}),
\end{align}
where $M_t:=B(\Sigma)^t(\Sigma_0-\Sigma)(B(\Sigma)^\top)^t$. We bound the last term 
in the above expression using the fact
\begin{align}
    \exists\ \epsilon\in(0,1):\ \tr\left(4\frac{1-\epsilon}{\epsilon^2}\Sigma-M_t\right) \ge 0\quad\forall\ t\ge 0.
\end{align}
This is easy to confirm since the range of the function 
$g(\epsilon)=4\frac{1-\epsilon}{\epsilon^2}$ in $(0,1)$ is $(0,\infty)$ 
and the matrix in the trace is symmetric. With this, we have
\begin{align}
    &\tr\left(4\frac{1-\epsilon}{\epsilon^2}\Sigma-M_t\right)\ge 0.\\
    \Leftrightarrow\ &
    \tr\left(4\frac{1-\epsilon}{\epsilon^2}\Sigma M_t-M_t^2\right)\ge 0.\\
    \Leftrightarrow\ &
    \tr\left((1-\epsilon)\Sigma M_t-\frac{\epsilon^2}{4}M_t^2\right)\ge 0.\\
    \Leftrightarrow\ &
    \tr\left(\Sigma M_t\right)\ge \tr\left(\epsilon\Sigma M_t+\frac{\epsilon^2}{4}M_t^2\right).\\
    \Leftrightarrow\ &
    \tr\left(\Sigma^2 + \Sigma M_t\right)\ge \tr\left(\Sigma^2+\epsilon\Sigma M_t+\frac{\epsilon^2}{4}M_t^2\right).\\
    \Leftrightarrow\ &
    \tr\left(\Sigma^2 + \Sigma^{1/2} M_t\Sigma^{1/2}\right)\ge \tr\left(\left(\Sigma + \frac{\epsilon}{2}M_t\right)^2\right).
\end{align}
Finally, since the square root operation is monotone increasing, we have
\begin{align}\label{M_t_bound}
    \tr\left((\Sigma^2 + \Sigma^{1/2} M_t\Sigma^{1/2})^{1/2}\right)\ge \tr(\Sigma)+ \frac{\epsilon}{2}\tr(M_t).
\end{align}
Plugging \eqref{M_t_bound} into \eqref{wass2_init} gives us
\begin{align}
    \Wass_2^2(\pi,\pi_t) &\le \|B(\Sigma)^t(\mu_0-\mu)\|_2^2 + 2\tr(\Sigma)+ \tr(M_t) - 2\tr(\Sigma) - \epsilon \tr(M_t)\\
    &=\|B(\Sigma)^t(\mu_0-\mu)\|_2^2 + (1-\epsilon)\tr(M_t)\\
    &=\|B(\Sigma)^t(\mu_0-\mu)\|_2^2 + (1-\epsilon)\tr(B(\Sigma)^t(\Sigma_0-\Sigma)(B(\Sigma)^\top)^t).
\end{align}
Combining this with Lemma 4 from \cite{roberts1997updating}, we have
\begin{align}
    \Wass_2^2(\pi,\pi_t) = O(\rho(B(\Sigma))^{2t})
\end{align}
which shows the convergence rate for Wasserstein 2 distance.
By properties of the Wasserstein distance, it follows that $\Wass_1(\pi_t, \pi) \leq \Wass_2(\pi_t, \pi)$ 
and hence $\Wass_1(\pi_t, \pi) = O(\rho(B(\Sigma))^t)$.

\paragraph{Chi-squared bound.}
For the Pearson-$\chi^2$ divergence, the result follows from direct application of 
Theorem 1 and Lemma 4 of \cite{li2005convergence}. Namely, we have that 
\[
\chi^2(\pi_t, \pi) = O(\rho(B(\Sigma))^t).
\]
\eprfof


\blem 
\label{lem:rho_B_Sigma}
Under the conditions of \cref{thm:Gibbs_convergence}, 
\[
  \rho(B(\Sigma)) \leq \exp\left(-\frac{1}{\kappa(\Sigma)}\right).
\]
\elem 

\bprfof{\cref{lem:rho_B_Sigma}}
Our proof proceeds in two steps, using results from \citet{roberts1997updating}.
We first establish a bound on the $L^2$ convergence rate of the random sweep Gibbs sampler (RSGS), 
$\rho_{L^2}(RSGS, \Sigma)$, in terms of $d$ and $\kappa(\Sigma)$. 
This rate is easier to study as it involves the matrix $A(\Sigma)$ instead of $B(\Sigma)$.
We then conclude that $\rho(B(\Sigma)) \leq \rho_{L^2}(RSGS, \Sigma)$ from an existing 
result in \citet{roberts1997updating}.

For any symmetric positive definite matrix $\Sigma$, there exists 
a rotation matrix $R$ that makes $R^{-1} \Sigma R$ diagonal. 
Furthermore, the condition number $\kappa(\Sigma)$ of the Gaussian distribution with covariance $\Sigma$ is invariant 
to rotations. Therefore, we can consider $\Norm(\mu,\Sigma)$ with a fixed condition 
number, \emph{diagonal} covariance matrix $\Sigma$ and its rotations without loss of 
generality. Now, for any matrix $M$, denote $\lambda(M)$ 
its set of eigenvalues and $D_M$ the inverse of the diagonal of $M$ if it has 
invertible diagonal elements. Let $\lambda_1\ge\ldots\lambda_d > 0$ 
denote the eigenvalues of $\Sigma$ so that $\kappa(\Sigma)=\lambda_1/\lambda_d$. 
Rotating $\Norm(\mu,\Sigma)$ using a rotation matrix $R$ gives us $\Norm(R\mu,R\Sigma R^\top)$.
The RSGS coordinate update matrix for $\Norm(R\mu,R\Sigma R^\top)$ is 
\[
    A(R\Sigma R^\top) = I-D_{RQR^\top}RQR^\top.
\]
To study the maximum eigenvalue of this matrix, we first note two matrix properties. 
First, for all symmetric positive definite matrices $M,N$, using the spectral norm, 
we have the inequality
\[
    \frac{1}{\min\lambda(MN)}
    = \|(MN)^{-1}\|
    \le\|M^{-1}\|\|N^{-1}\| 
    =\frac{1}{\min\lambda(M)}\frac{1}{\min\lambda(N)}.
\]
Also, since $\sum_{j=1}^{d}r^2_{i,j}=1$ for the matrix $R$, we have
\[
    \sum_{j=1}^{d}\frac{r^2_{i,j}}{\lambda_j}
    \geq \sum_{j=1}^{d}\frac{r^2_{i,j}}{\lambda_1}
    =\frac{1}{\lambda_1}.
\]
Combining these results, we have
\[
    \max \lambda(A(R\Sigma R^\top)) 
    &= 1-\min\lambda(D_{RQR^\top}RQR^\top)\\
    &\le  1-\min\lambda(D_{RQR^\top})\min\lambda(RQR^\top)\\
    &=1-\min\lambda(D_{RQR^\top})\min\lambda(Q)\\
    &=1-\min\lambda(D_{RQR^\top})\lambda_d\\
    &=1-\lambda_d\min_{i}\sum_{j=1}^{d}\frac{r^2_{i,j}}{\lambda_j}\\
    &=1-\frac{1}{\kappa(\Sigma)}.
\]
Now, let $\text{rot}_d$ be the set of $d$ by $d$ rotation matrices.
By Theorem 2 in \cite{roberts1997updating}, we have
\[ \label{eq:RSGS_bound}
    \rho_{L^2}(RSGS,\Sigma)
    &\leq \sup_{R\in\text{rot}_d}\rho_{L^2}(RSGS,R\Sigma R^\top) \\
    &=\sup_{R\in\text{rot}_d}\left(\frac{1}{d}(d-1+\max\lambda(A(R\Sigma R^\top)))\right)^d \\
    &\le \left(\frac{1}{d}\left(d-1+1-\frac{1}{\kappa(\Sigma)}\right)\right)^d \\
    &\leq \exp\left(-\frac{1}{\kappa(\Sigma)}\right).
\]

Finally, combining \cref{eq:TV_convergence}, \cref{eq:RSGS_bound} and Theorems 4 and 6 in 
\cite{roberts1997updating} gives us
\[
  \rho(B(\Sigma))
  = \rho_{L^2}(DUGS, \Sigma)
  \leq \rho_{L^2}(RSGS,\Sigma) 
  \leq \exp\left(-\frac{1}{\kappa(\Sigma)}\right).
\]
\eprfof


\subsection{Proof of \cref{prop:conditioning_invariance_Gibbs}}
Let $\Sigma' := D\Sigma D$ be the covariance matrix after diagonal preconditioning 
using the diagonal matrix $D$ and set $Q' = (\Sigma')^{-1} = D^{-1} \Sigma^{-1} D^{-1}$. 
\[
  A(\Sigma') 
  &= I - \diag((Q')_{11}^{-1}, (Q')_{22}^{-1}, \ldots, (Q')_{dd}^{-1})Q'\\
  &= DD^{-1} - D\diag(Q_{11}^{-1}, Q_{22}^{-1}, \ldots, Q_{dd}^{-1})D^{-1}DQD^{-1}\\
  &= D(I-\diag(Q_{11}^{-1}, Q_{22}^{-1}, \ldots, Q_{dd}^{-1})Q)D^{-1}\\
  &= DA(\Sigma)D^{-1}\\
  &= D(U(\Sigma)+L(\Sigma))D^{-1}\\
  &= DU(\Sigma)D^{-1}+DL(\Sigma)D^{-1}.
\]
Since $DU(\Sigma)D^{-1}$ and $DL(\Sigma)D^{-1}$ are upper and lower triangular 
matrices, respectively, we get
\[
  U(\Sigma')=DU(\Sigma)D^{-1},\quad L(\Sigma')=DL(\Sigma)D^{-1}.
\]
Next, we have
\[
  B(\Sigma')&=(I-L(\Sigma'))^{-1}U(\Sigma')\\
  &=(DD^{-1}-DL(\Sigma)D^{-1})^{-1}DU(\Sigma)D^{-1}\\
  &=D(I-L(\Sigma))^{-1}U(\Sigma)D^{-1}\\
  &=DB(\Sigma)D^{-1}.
\]
Now, let $\lambda$ and $x$ be an eigenvalue and eigenvector of $B(\Sigma')$, then
\[
  & B(\Sigma')x = \lambda x\\
  \Leftrightarrow & DB(\Sigma)D^{-1}x=\lambda x\\
  \Leftrightarrow & B(\Sigma)(D^{-1}x)=D^{-1}\lambda x.
\]
Therefore, $D^{-1}x$ is an eigenvector of $B(\Sigma)$ and $\lambda$ is an eigenvalue 
of $B(\Sigma)$ so that any eigenvalue of $B(\Sigma')$ is an eigenvalue of $B(\Sigma)$.
Similarly, we also have any eigenvalue of $B(\Sigma)$ is an eigenvalue 
of $B(\Sigma')$. Hence, $B(\Sigma)$ and $B(\Sigma')$ have the same set of 
eigenvalues which means they also have the same maximum eigenvalue modulus. 
That is, $\rho(B(\Sigma)) = \rho(B(\Sigma'))$.

\section{From theory to practice}\label{app:theory-practice}
In this section, we provide additional details on \cref{sec:theory} and make the connection
from theoretical results to experimental results.

\subsection{Divergences between distributions}\label{sec:divergences}
Let $\pi_1,\pi_2$ be given distributions and let $\mathcal{C}(\pi_1,\pi_2)$ be the set of all
couplings of $\pi_1$ and $\pi_2$ (joint distributions admitting $\pi_1$ and $\pi_2$ as marginals).
The divergences used to study the convergence of samplers to their target distributions
in most works include the total variation ($\TV$) distance
\[
	\label{eq:tv_distance}
	\TV(\pi_1, \pi_2) = \inf_{(X,Y)\in \mathcal{C}(\pi_1,\pi_2)} \P(X\neq Y),
\]
the $p$-Wasserstein ($\mathrm{W}_p$) ($p \geq 1$) distance
\[
	\label{eq:wp_distance}
	\mathrm{W}_p(\pi_1, \pi_2)
	= \left(\inf_{(X,Y)\in \mathcal{C}(\pi_1,\pi_2)} \E(\|X - Y\|^p)\right)^{1/p},
\]
and the Pearson-$\chi^2$ divergence
\[
	\label{eq:chi2_divergence}
	\chi^2(\pi_1, \pi_2)
	= \E\left[\left(\frac{\pi_1(X)-\pi_2(X)}{\pi_2(X)}\right)^2\right], \quad X\sim\pi_2.
\]
More generally, we can define the Pearson-$\chi^2$ divergence for any $\pi_1 \ll \pi_2$ as 
\[
  \chi^2(\pi_1, \pi_2)
  = \int \lt| \frac{\d \pi_1}{\d \pi_2} - 1\rt|^2 \, \d \pi_2.
\]

The divergences presented above have several properties. We state some here without proof. 
For instance, 
\[
  \mathrm{W}_p(\pi_1, \pi_2) \geq \mathrm{W}_q(\pi_1, \pi_2), \qquad p \geq q \geq 1,
\]
and 
\[
  \TV(\pi_1, \pi_2) \leq \frac{1}{2} \sqrt{\chi^2(\pi_1, \pi_2)}. 
\]

\subsection{Convergence rates and mixing times}\label{sec:rate_vs_mix_time}
Another quantity of interest is the mixing time of a sampler,
which is defined by the number of iterations required for a sampler to be 
within a certain divergence of its target distribution.
Precisely, the \emph{mixing time} of a Markov kernel $K(\cdot, \cdot)$ with initial
distribution $\mu_0$ and target distribution $\pi$, with respect to a divergence $\mathrm{D}$,
is defined as:
\[
	\label{eq:mixing_time}
	t_\text{mix}^K(\mathrm{D}, \epsilon, \mu_0, \pi)
	:= \inf\{t:\mathrm{D}(K^t\mu_0, \pi) \le \epsilon\},\quad \epsilon > 0.
\]
The \emph{mixing rate} is often used to compare the scaling of a certain quantity. 
Depending on which quantity that is of interest (e.g., $d$ or $\kappa$),
we can throw away low order terms in the mixing time with respect to that quantity.
For instance, a mixing time of $O(d \log d)$ correspond to mixing rate $O(d)$ if we focus on $d$ scaling.

An important distinction between convergence rate and mixing time is that convergence rate 
is asymptotic with respect to the number of iterations $t$, which does not capture its precise 
dependence on $d$ and $\kappa$. The main reason for this is due to the possible dependency 
on $d$ and $\kappa$ of the constant in front of the convergence rate. Therefore, a tight 
mixing time bound can be a more comprehensive metric of performance for a sampler of interest.

\subsection{From convergence rate to ESS} \label{sec:convrate_vs_ESS}

Throughout this paper, we have relied on the effective sample size (ESS) to 
evaluate the performance of different MCMC samplers. On the other hand, the 
theory we have discussed has mainly focused on mixing rates or times of MCMC
algorithms with respect to different divergences.
In this section, we provide two results that connect the theoretical convergence 
rate in total variation (TV) and $\chi^2$-divergence to the ESS.

Let us begin by introducing a few concepts.
We say that a $\pi$-invariant Markov kernel $K$ and test function $f\in L^2(\pi)$
satisfy a CLT if there exists $0<\sigma_f^2<\infty$ such that
\[\label{eq:def_CLT}
	\sqrt{T}\left( \frac{1}{T} \sum_{t=0}^{T-1} f(X_t) - \pi(f) \right) \convd \Norm(0, \sigma_f^2), \quad \text{as } T \to \infty.
\]
See \citet{roberts2004general} for a discussion of necessary and sufficient 
conditions. When \cref{eq:def_CLT} holds, the variance of the running average 
satisfies \citep[see e.g.][Prop. 1]{schmeiser1995optimal}
\[\label{eq:asymp_var_mean}
\Var\left(\frac{1}{T} \sum_{t=0}^{T-1} f(X_t)\right) = \frac{\sigma_f^2}{T} + O(T^{-2}).
\]
To see the importance of this equation, let us first simplify the notation 
(without loosing generality) by concentrating on the subset of zero-mean test 
functions $f \in L^{2}_0(\pi) := \{f\in L^2(\pi): \pi(f) = 0\}$.
Then \cref{eq:asymp_var_mean} can be rewritten as
\[\label{eq:asymp_var_mean_ESS}
\Var\left(\frac{1}{T} \sum_{t=0}^{T-1} f(X_t)\right) = \frac{\pi(f^2)}{\ESS(f)} + O(T^{-2}).
\]
where
\[
\ESS(f) := T \frac{\pi(f^2)}{\sigma_f^2},
\]
is the ESS of the test function $f$. \cref{eq:asymp_var_mean_ESS} shows that,
asymptotically in $T$, the variance of the running average $T^{-1}\sum_{t=0}^{T-1} 
f(X_t)$ is approximately equal to the variance achieved by a simple Monte Carlo 
estimator based on roughly $\ESS(f)$ \iid samples from $\pi$. In particular,
$\ESS(f)=T$ for every $f$ in the ideal case where $K$ produces exact \iid 
samples from $\pi$---that is, when $K(x,\cdot)=\pi$ for $\pi$-almost every $x$.

In the following we rely on the fact that, when \cref{eq:def_CLT} holds, 
$\sigma_f^2$ can be written as \citep{roberts2004general}
\[
\sigma_f^2 = \pi(f^2) + 2 \sum_{t=1}^\infty \Cov(f(X_0), f(X_t)),
\]
with $X_0\sim\pi$ and $X_{t+1}|X_t \sim K(X_t,\cdot)$ for all $t\geq0$.

\paragraph{Connection between TV and ESS}
The following result shows that for uniformly ergodic Markov chains with rate 
$\rho\in(0,1)$, the ESS of a certain class of functions admits a uniform lower
bound that is decreasing in $\rho$. In other words, a more efficient sampler---that
is, one with $\rho$ closer to $0$---should produce a uniformly higher ESS for a
certain class of test functions.

\bprop
\label{prop:tvESS}
Suppose that there exists a constant $C \geq 0$ and $\rho \in (0, 1)$ such that for 
$\pi$-almost every $x \in \scX$ we have
\[
	\TV(K^t(x, \cdot), \pi) \leq C \rho^t.
\]
Then, for any test function $f \in L_0^2(\pi)$ with $\abs{f} \leq 1$, we have that
\[
	\ESS(f) \geq \frac{T}{1 +  \frac{4C}{\pi(f^2)}\frac{\rho}{(1-\rho)}}.
\]
\eprop

\bprfof{\cref{prop:tvESS}}
Recall that the total variation distance between two probability measures
$\mu$ and $\nu$ 
can be characterized as \citep[see e.g.][Prop 3(b)]{roberts2004general}
\[
\TV(\mu, \nu) = \frac{1}{2}\sup_{|f|\leq 1} |\mu(f) - \nu(f)|,
\]
where the supremum is taken over all bounded functions $f:\mathcal{X}\to[-1,1]$. 
Hence, for any $f\in L_0^2(\pi)$ with $|f|\leq 1$ we have
\[
|K^t f(x)| \leq 2C\rho^t, \quad \pi-a.e.\ x.
\]
Then, using the Markov property, Jensen's inequality, and the fact that 
$|f|\leq 1$, we obtain 
\[
	\abs{\Cov(f(X_0), f(X_t))}
	= \abs{\E[f(X_0) f(X_t)]}
	= \abs{\pi(f K^t f)}
	\leq \pi(|f| |K^t f|)
	\leq \pi(|K^t f|)
	\leq 2C \rho^t.
\]
Consequently,
\[
\sum_{t=1}^\infty \pi(f K^t f) \leq \frac{2C\rho}{1-\rho},
\]
from which the result follows immediately.
\eprfof

Suppose now that we consider not one target distribution $\pi$ but a sequence $\pi_1, \pi_2, \dots$ of 
increasing complexity. For example, they might involve increasing dimensionality $d_i$ and/or 
condition number $\kappa_i$. In such case, if we have
\[
\TV(K^t(x, \cdot), \pi_i) \leq C \rho_i^t,
\]
then we can make predictions on the decay of the relative effective sample size as $i \to \infty$.
To do so in an interpretable way, we first rewrite $\rho_i$ as 
\[
\rho_i = 1 - \frac{1}{z_i},
\]
where $z_i \to \infty$ is a measure of complexity. For instance, suppose $z_i = i^{1/4}$ 
for optimally tuned and Metropolized HMC when targeting an $i$-dimensional isotropic 
normal distribution $\pi_i$.

\bcor
\label{cor:tvESS}
Suppose that there exists a constant $C \geq 0$ and $\rho_i \in (0, 1)$ such that for 
$\pi_i$-almost every $x \in \scX$ we have
\[
\TV(K^t(x, \cdot), \pi_i) \leq C \rho_i^t,
\]
where $\rho_i = 1 - \frac{1}{z_i}$ and $z_i \to \infty$. 
Then, for any sequence of test functions $\cbra{f_i}_{i \geq 1}$ with each
$f_i \in L_0^2(\pi_i)$, $\abs{f_i} \leq 1$, and $\limsup_{i \to \infty} \pi_i(f_i^2) < \infty$,
we have 
\[
  \text{relative ESS} = \frac{\ESS(f_i)}{T}  \geq b_i \sim \frac{\tilde C_i}{z_i}, \quad i\to\infty,
\]
where $\tilde C_i = \pi_i(f_i^2)/(4C)$.
\ecor

\bprfof{\cref{cor:tvESS}}
From \cref{prop:tvESS}, we can pick 
\[
b_i = \lt(1 +  \frac{4C}{\pi_i(f_i^2)}\frac{\rho_i}{(1-\rho_i)}\rt)^{-1}.
\]
From our assumptions on $z_i, \rho_i$, we have $\rho_i / (1-\rho_i) = (1 - 1/z_i) / (1/z_i) = z_i - 1$. 
Let $\tilde C_i = \pi_i(f_i^2) / (4C)$. We can rewrite the above expression as 
\[
  b_i = \lt(1 +  \frac{1}{\tilde C_i}(z_i - 1)\rt)^{-1}.
\] 
It remains to show that $b_i \sim \tilde C_i / z_i$ as $i\to \infty$: 
\[
  \lim_{i\to\infty} \frac{\tilde C_i / z_i}{b_i} 
  = \lim_{i\to\infty} \frac{\tilde C_i}{z_i} \left(1 +  \frac{1}{\tilde C_i}(z_i - 1)\right) 
  = \lim_{i\to\infty} \left( \frac{\tilde C_i}{z_i} + 1 - \frac{1}{z_i}\right) 
  = 1.
\]
\eprfof

For example, in the case of optimally tuned Metropolized HMC on isotropic normal targets, 
if $z_i = i^{1/4}$, 
we obtain from the above corollary that the relative ESS decays at the relatively slow rate of at most 
$1/z_i = i^{-1/4}$.

\paragraph{Connection between $\chi^2$ and ESS}
For a reversible Markov kernel $K$, \cref{prop:chiESS} below allows us to translate 
the convergence rate in $\chi$-squared divergence
into a uniform lower bound on the ESS of $L_2(\pi)$ functions.

We start with a key technical lemma that draws the connection between the convergence rate in $\chi$-squared divergence and the decay of variance for $f \in L_2^0(\pi)$.
\blem \label{lem:chitol2}
Suppose the Markov kernel $K$ is reversible with respect to $\pi$.
Further suppose that there exists $\rho \in [0, 1)$ such that for any initial distribution $\pi_0 \ll \pi$ satisfying $\chi^2(\pi_0, \pi) < \infty$ and for all $t \geq 0$, we have
\[ \label{eq:chiconv}
	\chi^2(\pi_0 K^t, \pi) \leq \rho^t \chi^2(\pi_0, \pi).
\]
Then, for all $f \in L_2^0(\pi)$ and for all $t >0$, we have
\[
    \|K^tf\|_{L_2(\pi)}^2 \leq \rho^t \|f\|_{L_2(\pi)}^2.
\]
\elem

\bprfof{\cref{lem:chitol2}}
First, by the reversibility of $K$ (and $K^t$ for all $t > 0$) with respect to $\pi$, 
we have that for all distributions $\pi_0$ such that $\chi^2(\pi_0, \pi) < \infty$, 
\[
    \label{eq:reversibility}
    K^t \frac{\d \pi_0}{\d \pi}  = \frac{\d  \pi_0 K^t}{\d \pi} \qquad \text{($\pi$-a.e.)}.
\]
To see this, we observe that for all $g \in L_2(\pi)$, 
\[
    \int K^t\left(\frac{\d \pi_0}{\d \pi}\right) g \d \pi 
    & = \int  \frac{\d \pi_0}{\d \pi} K^t(g) \d \pi \quad (\text{by the reversibility of } K^t)\\
    & = \int K^t g \d \pi_0\\
    & = \pi_0 K^t(g) \\
    & = \int \frac{\d  \pi_0 K^t}{\d \pi} g \d \pi.
\]
\cref{eq:reversibility} combined with the definition of the $\chi$-squared divergence then allows us to interpret 
$\chi^2(\pi_0 K^t, \pi) \leq \rho^t \chi^2(\pi_0, \pi)$ as 
\[\label{eq:rewritechiconv}
   \left\|K^t \left( \frac{\d \pi_0}{\d \pi} - 1\right)\right\|_{L_2(\pi)}^2 \leq \rho^t \left\|\frac{\d \pi_0}{\d \pi} - 1\right\|_{L_2(\pi)}^2.
\]
Under the assumption of \cref{lem:chitol2}, we have that \cref{eq:rewritechiconv} 
holds for all $t$ and for all $\pi_0 \ll \pi$ such that $\chi^2(\pi_0, \pi) < \infty$.

Consequently, we can conclude that for all bounded $f \in L_2^0(\pi)$ 
(i.e., $\sup_x |f(x)| < \infty$) and for all $t >0$, we have
\[
\label{eq:boundedfunction}
  \|K^tf\|_{L_2(\pi)}^2 \leq \rho^t \|f\|_{L_2(\pi)}^2.
\]
To see this, define the measure 
\[
  p_f(A) = \int_A \frac{f}{\sup_x |f(x)|} + 1 \, \d \pi, 
\]
which satisfies $p_f \ll \pi$ and $\chi^2(p_f, \pi) < \infty$. 
Then, 
\[ 
\frac{\d p_f}{\d \pi} = \frac{f}{\sup_x |f(x)|} + 1,
\]
and applying \cref{eq:rewritechiconv} with $p_f$ in place of $\pi_0$, we get \cref{eq:boundedfunction}.

Finally, we focus on removing the boundedness assumption on $f$. 
For any $f \in L_2^0(\pi)$, define 
\[
f_{L, M}(x) := \left\{\begin{array}{ll}
        \min\{f(x), M\} & \text{as } x \geq 0 \\
        \max\{f(x), -L\} & \text{as } x < 0 \\
\end{array} 
\right.
\]
Notice that for any $M > 0$, there exists $L > 0$ such that $\pi(f_{L, M}) = 0$.

Consider an increasing sequence $M_n$ such that $\lim_{n \to \infty} M_n = \infty$.
For any $f \in L_2^0(\pi)$, we can construct an approximating sequence $f_n := f_{L_n, M_n}$, where $L_n$ is chosen so that
for all $n \in \nats$, $f_{L_n, M_n} \in L_2^0(\pi)$. 
Notably, $L_n$ has to be a non-decreasing sequence, $\lim_{n\to \infty} \|f_n -f\|_{L_2(\pi)}= 0$,
and $\sup_{x} f_{L_n, M_n} = \max\{L_n, M_n\} < \infty$.
Therefore, for all $f\in L_2^0(\pi)$ and for all $n \in \nats$, we have that 
\[\label{eq:boundedseq}
    \|K^tf_n\|_{L_2(\pi)}^2 \leq \rho^t \|f_n\|_{L_2(\pi)}^2, \quad \text{ for all } t > 0.
\]
Because $\|f_n\|_{L_2(\pi)}$ converges increasingly to $\|f\|_{L_2(\pi)}$
as $n\to \infty$, we invoke the monotone convergence theorem to take the limit
$n \to \infty$ on both sides of \cref{eq:boundedseq}, which completes the proof.
\eprfof

\bprop
\label{prop:chiESS}
Under the same conditions of \cref{lem:chitol2}, we have that 
\[
	\inf_{f\in L_0^2(\pi)} \ESS(f) \geq  \frac{T}{1 + 2\frac{\sqrt{\rho}}{1-\sqrt{\rho}}}.
\]
\eprop

\bprfof{\cref{prop:chiESS}}
\cref{lem:chitol2} yields that for all $f \in L_2^0(\pi)$ and all 
$t \geq 0$,
\[
  \pi((K^tf)^2) \leq \rho^t \pi(f^2).
\]
Therefore, by the Cauchy--Schwarz inequality,
\[
  |\pi(f K^t f)| \leq \sqrt{\pi(f^2)}\sqrt{\pi((K^tf)^2)} \leq \pi(f^2)(\sqrt{\rho})^t.
\]
Hence,
\[
  \sum_{t=1}^\infty \pi(f K^t f) \leq \pi(f^2)\frac{\sqrt{\rho}}{1-\sqrt{\rho}},
\]
and the claim then follows by replacing the above in the definition of $\ESS(f)$
and then taking the infimum.
\eprfof

We provide an analogous result to \cref{cor:tvESS} for a reversible kernel $K$ 
in terms of $\chi$-squared divergence convergence.

\bcor
\label{cor:chiESS}
Consider a sequence of target distributions $\pi_i$.
Suppose there exists $\rho_i \in (0, 1)$ such that for all $i$, for all $t > 0$, and  
for all initial distributions $\mu_{0,i}$ satisfying $\chi^2(\mu_{0,i}, \pi_i) < \infty$,
we have
\[ 
  \chi^2(\mu_{0,i} K^t, \pi_i) \leq \rho_i^t \chi^2(\mu_{0,i}, \pi_i),  
\]
where $\rho_i = 1 - \frac{1}{z_i}$ and $z_i \to \infty$. 
Then, for any sequence of test functions $\cbra{f_i}_{i \geq 1}$ with each
$f_i \in L_0^2(\pi_i)$, we have 
\[
\text{relative ESS} = \frac{\ESS(f_i)}{T}  \geq \frac{1}{1 + 2\frac{\sqrt{\rho_i}}{1-\sqrt{\rho_i}}}
 \sim \frac{1}{4 z_i}, \; i\to\infty.
\]
\ecor

\bprfof{\cref{cor:chiESS}}
We only need to show that 
$\frac{1}{1 + 2\frac{\sqrt{\rho_i}}{1-\sqrt{\rho_i}}} \sim \frac{1}{4 z_i}$.
We do this by showing that 
\[
  \frac{\sqrt{\rho_i}}{1-\sqrt{\rho_i}} \sim \frac{1}{1-\sqrt{\rho_i}} \sim 2 z_i.
\]
Because $\rho_i = 1-1/z_i$ and $z_i \to \infty$, we have $\rho_i \to 1$ and so the first asymptotic equivalence follows. 
Next, by a Taylor expansion of $h(x) = \sqrt{1-1/x}$ about $x = \infty$, we have  
$h(x) = 1-1/(2x) + o(1/x)$. Therefore, $1-\sqrt{\rho_i} \to 1/(2 z_i)$.
\eprfof

\begin{remark}
Note that the bounds in \cref{prop:tvESS,prop:chiESS} should be reasonably tight 
in the limit $\rho\to0$, since for $\rho=0$ they both recover the \iid case $\ESS(f)=T$.
\end{remark}

\subsection{From Gibbs to Metropolis-within-Gibbs}\label{sec:within-Gibbs}

While the theoretical results discussed in this section concern Gibbs samplers (i.e., using full
conditional updates),
the methodology described in Section~\ref{sec:fast_gibbs} applies to any ``Metropolis-within-Gibbs''
algorithm.
The reason for this gap is that the theoretical analysis of the Gibbs sampler
is more developed than that of Metropolis-within-Gibbs samplers.
Fortunately, recent work has shown how to relate the convergence properties of Metropolis-within-Gibbs
samplers with their idealized, full conditional counterparts \citep{ascolani2024gibbs}.
Invariance under axis-aligned stretching does not hold for Metropolis-within-Gibbs samplers.
However, algorithms such as the slice sampler with doubling are expected to be relatively insensitive to axis-aligned
transformations \citep{neal_slice_2003}.

\section{Connection between graphical model and compute graph}\label{app:cg_gm}
In this section, we sketch a simple proof that all savings made from graphical models 
derived Markov blankets are recovered as special cases of the cached compute graph method.

Consider a target distribution $\pi$ in $\mathbb{R}^d$ and, for any node $\theta_i$ 
in the graphical model of $\pi$, denote  the 
parent nodes of $\theta_i$ by $\mathcal{P}_i(\theta)$. Then, for all $\theta = (\theta_1,\ldots,\theta_d)\in \mathbb{R}^d$,
\[
  \pi(\theta) &= \prod_{i=1}^d p_i(\theta_i|\mathcal{P}_i(\theta)). 
\]
Now rewriting $p_i(\theta_i|\mathcal{P}_i(\theta)) = \exp f_i(S_i)$ where $S_i$ is the concatenation of $\theta_i$ with  $\mathcal{P}_i(\theta)$, we have
\[
  \log \pi(\theta) &= \sum_{i=1}^d f_i(S_i). 
\]
This last expression encodes the factor graph associated with the directed graphical model. 
To construct the compute graph of $\log \pi$ from its factor graph, we create a new node corresponding 
to the output $\log \pi(\theta)$ and connect all $f_i(S_i)$ nodes to this new node via a summation node. Crucially, that summation node is cached in the same way as described in Section~\ref{sec:fast_gibbs}, i.e., by subtracting and adding updated values. Finally, by expanding the nodes $f_i(S_i)$ 
nodes into their respective compute graphs, we obtain the compute graph for $\log \pi$.

Notice that when a node $\theta_i$ is updated, only the $f_j$'s connected to $\theta_i$ need to be updated. This set of $f_j$'s coincide with the Markov blanket. Moreover, the number of updates to the summation node is equal to the number of $f_j$'s updated.

\section{Effects of Gibbs sweep}\label{app:sweep-type}

\begin{figure}[t]
    \centering
    \begin{subfigure}{0.48\textwidth}
      \centering
      \includegraphics[width=\textwidth]{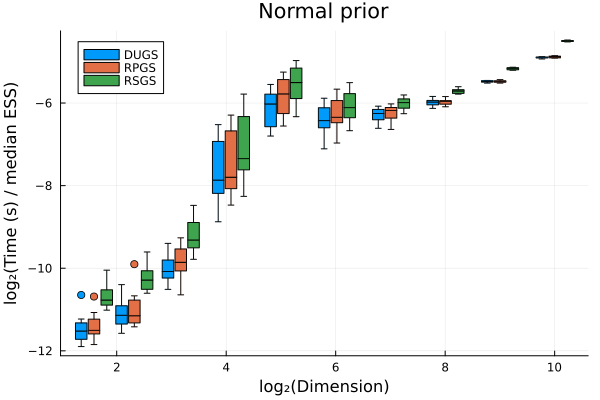}
    \end{subfigure}
    \begin{subfigure}{0.48\textwidth}
        \centering
        \includegraphics[width=\textwidth]{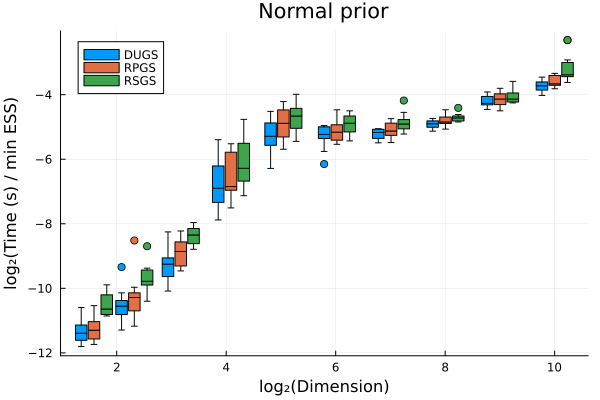}
    \end{subfigure}
    \caption{Time per ESS for CGGibbs a function of dimension for colon dataset with normal priors.
    The box plots summarize the results over 10 replicates.}
    \label{fig:dim_scale_Gibbs_sweep_normal}
\end{figure}

\begin{figure}[t]
    \centering
    \begin{subfigure}{0.48\textwidth}
      \centering
      \includegraphics[width=\textwidth]{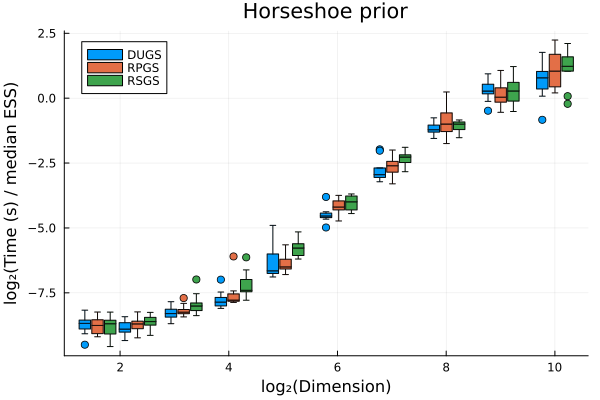}
    \end{subfigure}
    \begin{subfigure}{0.48\textwidth}
        \centering
        \includegraphics[width=\textwidth]{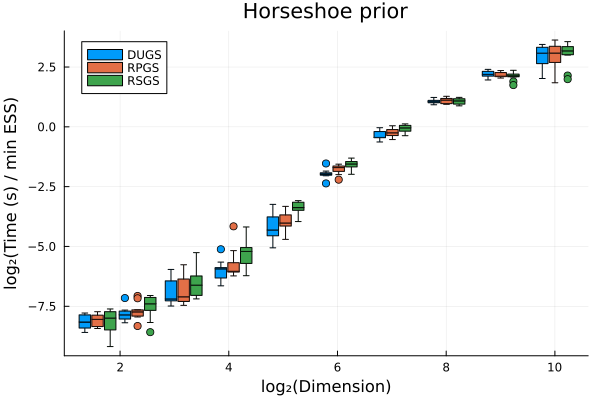}
    \end{subfigure}
    \caption{Time per ESS for CGGibbs a function of dimension for colon dataset with horseshoe priors.
    The box plots summarize the results over 10 replicates.}
    \label{fig:dim_scale_Gibbs_sweep_HSP}
\end{figure}

In this section, we investigate how different types of Gibbs samplers 
affect the time per ESS of the chain. Here, we consider 3 types of 
Gibbs samplers: DUGS, RSGS and random permutation Gibbs sampler (RPGS), 
where a random but fixed update order is generated each sweep. Next, we 
repeat the experiment in \cref{fig:dim_scale_colon} with both normal and 
horseshoe priors for the 3 Gibbs samplers. The results are shown in 
\cref{fig:dim_scale_Gibbs_sweep_normal} and \cref{fig:dim_scale_Gibbs_sweep_HSP}. 
We can see from these figures that the scalings in time per ESS for all 3 Gibbs samplers 
are relatively similar. This justifies our decision to only use DUGS as the representative 
for Gibbs samplers. Moreover, DUGS tends to have the best time per ESS, 
RSGS the worst and RPGS somewhere in between, further justifying our choice to use DUGS. 
This is somewhat expected since, for normal targets, we often have 
$\rho_{L^2}(DUGS)\le \rho_{L^2}(RSGS)$ \citep{roberts1997updating} and RPGS is intuitively 
a mix of DUGS and RSGS.

\end{document}